\documentclass[11pt,a4paper]{article}
\usepackage{jheppub}
\usepackage{epsfig}
\usepackage{amsmath}
\usepackage{subfigure}
\usepackage{graphicx}

\def \abs#1{\left\vert#1\right\vert}
\newcommand{\be}{\begin{equation}}
\newcommand{\ee}{\end{equation}}
\newcommand{\bea}{\begin{eqnarray}}
\newcommand{\eea}{\end{eqnarray}}
\newcommand{\bwt}{\begin{widetext}}
\newcommand{\ewt}{\end{widetext}}

\title{ A Singlet Extension of the MSSM with a Dark Matter Portal}

\author[a]{Alejandro de La Puente}
\author[b,c]{and Walter Tangarife}

\affiliation[a]{Theory Group, TRIUMF, Vancouver BC V6T 2A3, Canada}
\affiliation[b]{Theory Group, Department of Physics, The University of Texas, Austin, TX 78712, USA}
\affiliation[c]{Texas Cosmology Center, The University of Texas, Austin, TX 78712, USA}

\emailAdd{adelapue@triumf.ca}
\emailAdd{wtang@physics.utexas.edu}

\abstract{The minimal extension of the MSSM (NMSSM) has been widely studied in the search for a natural solution to
the $\mu$-problem. In this work, we consider a variation of the NMSSM where an additional singlet is added and a Peccei-Quinn symmetry is imposed. We study its neutralino sector and compute the annihilation cross section of the lightest neutralino. We use existent cosmological and collider data to constrain the parameter space and consider the lightest neutralino, with mass between 1 and 15 GeV, as a thermal dark matter candidate.}

\subheader{\begin{flushright}
UTTG-16-13\\
TCC-013-13
\end{flushright}}

\begin{document}
\maketitle

%%%%%%%%%%%%%%%%%%%%%%%%%%%%%%%%%%%%%%%%%%%%%%%%%%%%%%%%%%%%%%%%%%%%%%%%%%%%

%          Table of contents automatic !!!                                 %

%%%%%%%%%%%%%%%%%%%%%%%%%%%%%%%%%%%%%%%%%%%%%%%%%%%%%%%%%%%%%%%%%%%%%%%%%%%%
%\section{\bf Introduction}
\section{Introduction}

The long standing problem of evading the LEP bound~\cite{Barate:2003sz} on the mass of the lightest (Standard Model (SM)-like) Higgs boson within supersymmetric extensions of the SM (SUSY) has to now be reinterpreted due to the recent discovery of what appears to be a SM-like Higgs boson with a mass of $125$ GeV~\cite{Aad:2012tfa,Chatrchyan:2013lba}. However, the underlying nature of this problem remains the same. In the Minimal SUSY SM (MSSM) the lightest Higgs must lie below the $Z$ mass at tree level. This mass can be pushed up through radiative corrections arising from the third family of quarks and squarks. However, the discovery of a $125$ GeV SM-like Higgs boson has placed the MSSM into a region of parameter space where the hierarchy problem, which SUSY is expected to solve, is reintroduced. That is, reaching a value of $125$ GeV requires large stop masses, as heavy as $10$ TeV, or a tuned value of the stop mixing parameter at the electroweak scale~\cite{Draper:2011aa,Carena:2011aa}. This version of the original hierarchy problem is well known as the ``little hierarchy problem" and it is quite generic within the MSSM (see ~\cite{Dimopoulos:1995mi, Barbieri:2000gf,Kitano:2006gv} and references therein). 
{\color{black}

One popular route that is taken to alleviate this problem is to extend the Higgs sector of the MSSM. This has the effect of generating new quartic terms in the scalar potential~\cite{Espinosa:1992hp,Kane:1992kq}. These new quartic terms push up the mass of the SM-like Higgs boson at tree level, removing the need for large radiative corrections. In particular, this can be achieved by extending the MSSM sector with SM gauge singlets. The minimal extension of the MSSM, refered to as the next-to-minimal SUSY Standard Model or NMSSM, incorporates a single gauge singlet and was introduced primarily to address the $\mu$-problem of the MSSM (For reviews, see~\cite{Maniatis:2009re,Ellwanger:2008py,Martin:1997ns}.) This model has had its fair share of success, but it is not clear how one can naturally generate the Higgs mass on the order of $125$ GeV without introducing some degree of fine tuning, arguably as large as in the MSSM~\cite{Ellwanger:2006rm}. Furthermore, its minimal incarnation may introduce tension between the way the hierarchies are stabilized and the generation of domain walls~\cite{Abel:1995wk}. However, it has been shown that a stable NMSSM without domain walls is possible if certain discrete $R$-symmetries are imposed~\cite{Panagiotakopoulos:1998yw,Panagiotakopoulos:1999ah}. Nonetheless, the discovery of a SM-like Higgs boson has led to many studies on the phenomenology of a $125$ GeV SM-like Higgs boson within the NMSSM~\cite{King:2012is,Gunion:2012zd,Ellwanger:2012ke,Vasquez:2012hn,Gunion:2012gc,Bae:2012am,Agashe125gev,Badziak:2013bda,Badziak:2013gla}. In addition, it has been noted that a generalized version of the NMSSM that follows from underlying discrete $R$-symmetries can reduce the amount of fine tuning in the scalar sector~\cite{Ross:2011xv,Ross:2012nr,Kaminska:2013mya}. Within this class of models, additional operators are generated when SUSY and the $R$-symmetry are broken in the hidden sector and the effects are mediated to the observable sector through Planck-suppressed operators. Alternatively, one may generalize the MSSM by introducing effective dimension four and five operators~\cite{Dine:2007xi,Cassel:2009ps,Carena:2010cs}. These operators can reduce the amount of fine tuning in the scalar sector and be sensitive to new degrees of freedom at the TeV scale.

One particular generalization of the NMSSM, the S-MSSM~\cite{Delgado:2010uj,Delgado:2012yd}, has been implemented to fully address the little hierarchy problem of the MSSM. Within this framework one gives up any attempt at addressing the origin of the $\mu$-term and the absence of a $Z_{3}$ symmetry and incorporates a supersymmetric mass for the SM singlet field which is used to stabilize the singlet's vacuum expectation value ($vev$). A version of this model has been successfully embedded into a model where SUSY breaking is mediated by gauge interactions~\cite{Delgado:2010cw}.

It may be possible to argue that a successful natural solution to the little hierarchy problem and the $\mu$-problem exists within one unified model. In the scenario described in~\cite{Delgado:2012yd}, it was shown that one can successfully eliminate the $\mu$-term as a phenomenological parameter as long as a small supersymmetric mass for the singlet is incorporated. The model is described by the following superpotential:
\begin{equation}
W=W_{Yukawa}+\lambda\hat{S}\hat{H}_{u}\cdot\hat{H}_{d}+\frac{\mu_{S}}{2}\hat{S}^{2}.
\end{equation}
Of course, the above superpotential reintroduces a $\mu$-problem, a $\mu_{S}$-problem. The model we propose in this work goes one step further as it replaces the $\mu_{S}$-parameter by a second SM gauge singlet superfield, $\mu_{S}\to \rho \hat{N}$, with $\rho$ a dimensionless parameter. Within this framework the gauge singlet $\hat{N}$ has no direct couplings to the MSSM fields. In this way the S-MSSM gauge singlet, $\hat{S}$, serves as a portal for the singlet $\hat{N}$. This is possible due to the existence of a  (PQ) symmetry \footnote{For examples of previous works where a PQ symmetry is considered within the NMSSM scenario, see \cite{Jeong:2011jk, Bae:2012am}}.

Additionally, we explore the existence of a dark matter candidate within the model presented in this work. Currently, there is plenty of evidence that points towards the existence of dark matter (DM) in our universe~\cite{Bergstrom:2000pn,Bertone:2004pz}, providing strong evidence for physics beyond the Standard Model. Recent results from Planck~\cite{Ade:2013lta} suggest a cold dark matter component with a density of $\Omega h^{2}=0.1199\pm0.0027$. Furthermore, there exist positive signals in direct detection experiments \cite{Aalseth:2010vx,Agnese:2013rvf} that point to a light dark matter candidate with mass at around $10$ GeV. Commonly, supersymmetric models contain a light degree of freedom that is cosmologically stable. In fact, the lightest neutralino in our model has a mass between $1-15$ GeV and annihilates into SM particles, mainly leptons and light quarks, through the exchange of light CP-even and -odd Higgs bosons and neutral gauge bosons. We take a look at the regions of the parameter space that allow a relic density that agrees with observations and consistent with collider searches and constraints arising from Big Bang nucleosynthesis. 

The structure of this paper is as follows: In Section~\ref{sec:TheModel} we introduce the model and look at the structure of Electroweak Symmetry Breaking (EWSB). In Section~\ref{sec:Constraints} we review the constraints arising from colliders that limit our parameter space while in Section~\ref{sec:DarkMatter} we show the main annihilation channels contributing to the density of dark matter in the universe. In Section~\ref{sec:Conclusions}, we offer concluding remarks on the possibility of a light neutralino in singlet extensions of the SM.

%%%%%%%%%%%%%%%%%%%%%%%%%%%%%%%%%%%%%%%%%%%%%%%%%%%%%%%%%%%%%%%%%Peccei-Quinn
%
%
%
%
%%%%%%%%%%%%%%%%%%%%%%%%%%%%%%%%%%%%%%%%%%%%%%%%%%%%%%%%%%%%%%%%%

\section{Model}\label{sec:TheModel}
\subsection{Electroweak Symmetry Breaking and Scalar Higgs Sector}\label{sec:HiggsSec}

In this work, we modify the S-MSSM~\cite{Delgado:2010uj,Delgado:2012yd} by replacing the supersymmetric mass term for the SM gauge singlet $\hat{S}$ by an additional SM gauge singlet superfield $\hat{N}$. This new superfield does not couple directly to the fields in the MSSM, but only through the mixing induced by a superpotential coupling between $\hat{N}$ and $\hat{S}$. Additionally, we impose a Peccei-Quinn (PQ) symmetry where both MSSM Higgs doublets, $H_{u}$ and $H_{d}$, have charge 1 and the singlets $S$ and $N$ are given charges $-2$ and $4$ respectively. Furthermore, under this symmetry, quarks and leptons have $PQ$ charges of $-1/2$. Using this framework, the superpotential we consider, defined at some high energy scale, is given by:
\begin{equation}
W=W_{Yukawa}+\lambda\hat{S}\hat{H}_{u}\cdot\hat{H}_{d}+\rho\hat{N}\hat{S}^{2}+\frac{\kappa}{3}\hat{S}^{3},\label{eq:SupModel}
\end{equation}
where the $\kappa$ term is introduced to give mass to a Nambu-Goldstone boson that arises from the spontaneous breaking of the PQ symmetry due to EWSB. Additionally, SUSY breaking generates the following contributions to the scalar potential
\begin{eqnarray}
V_{Soft}&=&V_{Soft,Yukawa}+m^{2}_{H_{u}}|H_{u}|^{2}+m^{2}_{H_{d}}|H_{d}|^{2}+m^2_{S}|S|^{2}+m^2_{N}|N|^{2} \nonumber \\
&+&\left(\lambda A_{\lambda}SH_{u}\cdot H_{d}+\rho A_{\rho}NS^{2}+\frac{\kappa}{3}A_{\kappa}S^{3}+c.c \right),\label{eq:SoftModel}
\end{eqnarray}
where $H_{u}=(H^{+}_{u},H^{0}_{u})$ and $H_{d}=(H^{0}_{d},H^{-}_{d})$.

In this analysis, we are interested in the limit where $S$ and $N$ interact weakly, that is $\rho\ll 1.$ Furthermore, we consider only small values for $\kappa$ such that the PQ symmetry is only slightly broken. Although there is no symmetry that forbids the PQ symmetry breaking operators $\alpha_{1}\hat{N}^{2}\hat{S}$ and $(\alpha_{2}/3)\hat{N}^{3}$ and $\lambda_{N}\hat{N}\hat{H}_{u}\cdot\hat{H}_{d}$, we have set them to zero at the messenger scale. This is a scale-dependent assumption. Nevertheless, this hypothesis is quite stable under renormalization group effects. Indeed, assuming small values for $\alpha_{1}$, $\alpha_{2}$ and $\lambda_{N}$ at the messenger scale, these couplings remain small and well below $\rho$ and $\kappa$ at the electroweak scale. The running between a messenger scale given by $M_{mess}=10^{12}$ GeV and the weak scale is shown in Figure~\ref{fig:rge_plot}, where we have used the one-loop renormalization group equations for the dimensionless couplings given in Appendix~\ref{sec:rges}. Again, it is important to emphasize that a particular high energy choice for the superpotential in Equation~(\ref{eq:SupModel}) was made. The structure may be achieved with additional dynamics above the messenger scale. 

\begin{figure}[ht]\centering
\bigskip
\vspace{1cm}
\includegraphics[width=4.0in]{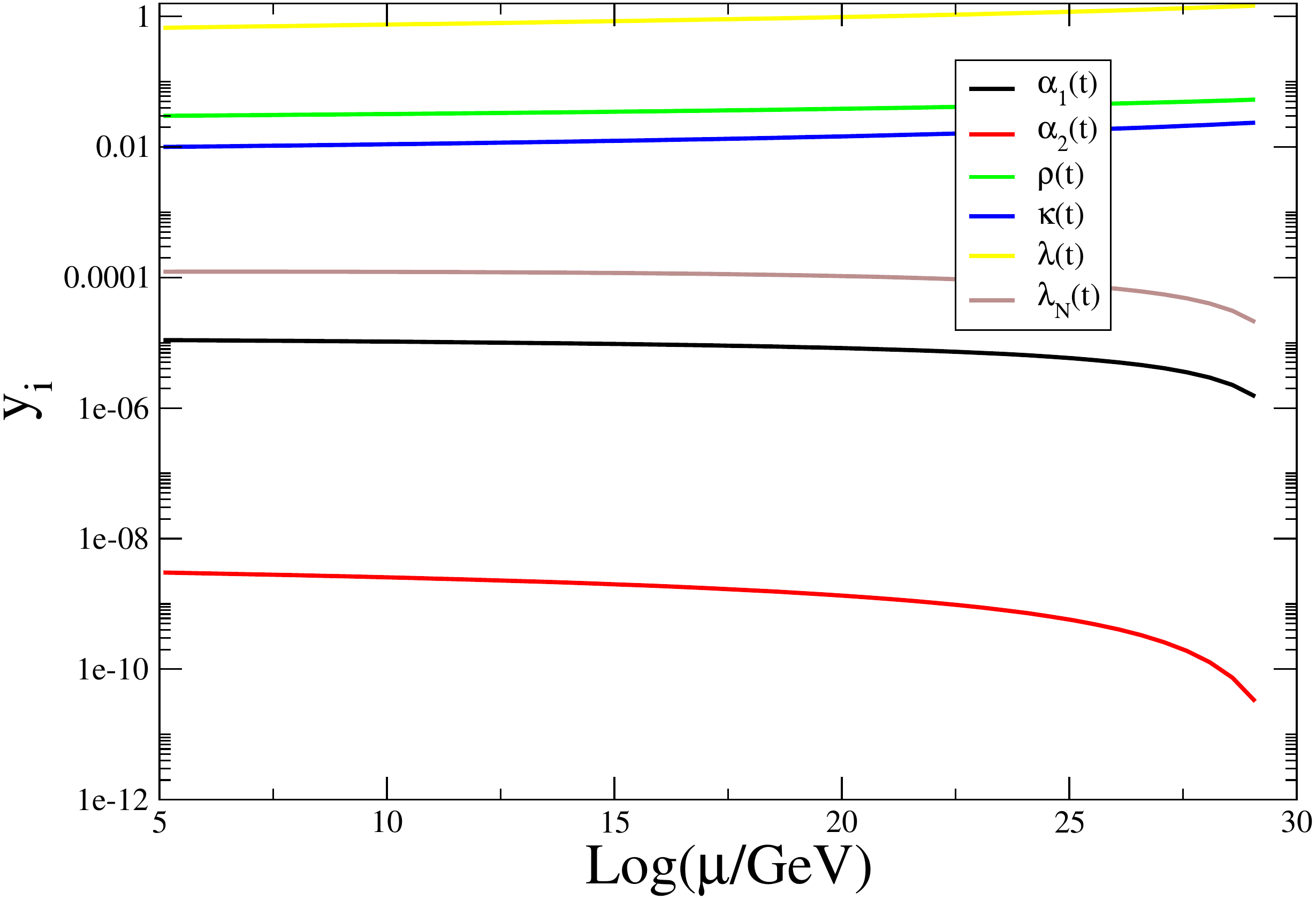}
\caption{\small Dimensionless couplings as function of $t=\log{\mu/\text{GeV}}$. The figure was generated by running the couplings between the electroweak scale, $M_{EW}=160$ GeV and a messenger scale given by  $M_{mess}=10^{12}$ GeV.} \label{fig:rge_plot}
\end{figure}

Furthermore, we note that once the PQ symmetry is broken by the $\kappa \hat{S}^{3}$ operator in Equation~(\ref{eq:SupModel}), additional contributions to $V_{soft}$ breaking this symmetry will be generated at one and two loops. These operators are further suppressed by powers of $\kappa,~\rho$ and $\lambda$. In addition, our model has an exact $Z_{3}$ symmetry as in the NMSSM. This discrete symmetry can lead to the generation of domain walls. However, our framework (with $\kappa\to 0$) has a $U(1)_{R}$ symmetry, as the one discussed in~\cite{Panagiotakopoulos:1999ah}, which has a $Z_{5}$ subgroup that can induce a tadpole term in the scalar potential large enough to avoid a domain wall problem without destabilizing the electroweak hierarchy. Nevertheless, our model contains two singlets and a more detailed analysis of the $R$-symmetries and their discrete subgroups is imperative for an in depth study of Equation~(\ref{eq:SupModel}). Suffice it to say that one may completely avoid an explicit PQ breaking term in the superpotential of Equation~(\ref{eq:SupModel}) and find an appropriate discrete $R$-symmetry that can induce tadpole terms capable of stabilizing the hierarchy and avoid the cosmological domain wall problem. 

The superpotential in Equation~(\ref{eq:SupModel}) together with the soft-breaking terms in Equation~(\ref{eq:SoftModel}) give rise to the following scalar potential for the neutral components of the two electroweak Higgs doublets and singlet fields:
\begin{eqnarray}
V^{0}_{H}&=&V_{soft}+\frac{1}{8}(g^{2}+g'^{2})(|H^0_{u}|^{2}-|H^0_{d}|^{2})^{2} +|S|^{2}\left(\lambda^{2}\left(|H^{0}_{u}|^{2}+|H^{0}_{d}|^{2}\right)+\rho^{2}|S|^{2}\right) \nonumber \\
&+&\left(\kappa S^{\dagger\,2}+\lambda H^{0}_{u}H^{0}_{d}+2\rho N^{\dagger}S^{\dagger}\right)\left(\kappa S^{2}+\lambda H^{0\dagger}_{u}H^{0\dagger}_{d}+2\rho NS\right)  
,\label{eq:PQNMSSMP}
\end{eqnarray}
where $g$ and $g'$ are the gauge couplings of the $SU(2)_{W}$ and $U(1)_{Y}$ gauge groups respectively. Minimizing the scalar potential with respect to $H^{0}_{u},~H^{0}_{d},~S$ and $N$ leads to the following four constraints:
\begin{eqnarray}
\sin2\beta&=&\frac{f(v_{S},v_{N})}{m^{2}_{H_{u}}+m^{2}_{H_{d}}+2\lambda^{2}v^{2}_{S}+\lambda^{2}v^{2}}, \nonumber \\ 
&\text{with}&~~~~~f(v_{S},v_{N})=2\left(\lambda v_{S}A_{\lambda}+2\lambda\kappa v^{2}_{S}+4\lambda\rho v_{S}v_{N}\right), \label{eq:1MinCon} \\
\frac{m^{2}_{Z}}{2}&=&\frac{m_{H^{2}_{d}}-m^{2}_{H_{u}}\tan^{2}\beta}{\tan^{2}\beta-1}-\lambda^{2}v^{2}_{S}, \label{eq:2MinCon} \\
v_{S}&=&\frac{\lambda A_{\lambda}v^{2}\cos\beta\sin\beta}{m^{2}_{S}+\lambda^{2}v^{2}} \label{eq:3MinCon}, \label{eq:3MinCon} \\
v_{N}&=&\frac{\left(\lambda v^{2}v_{S}\sin2\beta-2\kappa v^{3}_{S}- A_{\rho}v^{2}_{S}\right)}{m^{2}_{N}+4\rho^{2}v^{2}_{S}}\rho,\label{eq:4MinCon}
\end{eqnarray}
where $v_{S}=\left<S\right>$, $v_{N}=\left<N\right>$ and $v_{u,d}=\left<H_{u,d}\right>$ with $v=\sqrt{v^{2}_{u}+v^{2}_{d}}=174~\text{GeV}$ and $\tan\beta=v_{u}/v_{d}$. The vacuum expectation values for the two singlets have been obtained in the limits where both $\kappa$ and $\rho$ are much smaller than one. Equations~(\ref{eq:1MinCon}) and (\ref{eq:2MinCon}) are analogous to the MSSM minimization conditions with an effective $\mu$-parameter given by $\mu_{eff}=\lambda v_{S}$ and an effective $B_{\mu}$ term given by $B_{\mu,eff}=f(v_{S},v_{N})/2$.

In the absence of explicit CP-violating phases in the Higgs sector, the physical spectrum of the model includes a single charged Higgs boson $\left(H^{\pm}\right)$, four neutral scalars that we label $\left(h_{N},h_{S},h^{0},H\right)$ and three neutral pseudoscalars $\left(A_{N},A_{S},A\right)$. The states labeled with a subscript will turn out to have a large singlet component. For the state most resembling the usual pseudoscalar Higgs of the MSSM, the mass is given by
\begin{equation}
m^{2}_{A}\approx\frac{2B_{\mu,eff}}{\sin2\beta}.\label{eq:MSSMps}
\end{equation}
In the $\rho\to 0$ limit, $\rho\cdot v_{N}$ is largely suppressed and the effective supersymmetric mass for the singlet $S$, $\mu_{S,eff}$, is small. This is interesting since the spectrum of $H_{u},~H_{d}$ and $S$ mimics the one studied in~\cite{Delgado:2012yd}. In the analysis, in the limit where $\mu^{2}_{S}$, $m^{2}_{S}\ll \lambda^{2}v^{2}$, mixing of the singlet into the light MSSM-like scalar vanishes, yet receives an NMSSM-like enhancement~\cite{Ellwanger:2008py}:
\begin{equation}
m^{2}_{h^{0}}\approx m^{2}_{Z}\cos^{2}2\beta+\lambda^{2}v^{2}\sin^{2}2\beta-\frac{\left(m^{2}_{Z}-\lambda^{2}v^{2}\right)^{2}}{m^{2}_{A}}\sin^{2}2\beta\cos^{2}2\beta.\label{eq:SMSSMlightHiggs}
\end{equation}
Therefore, in the $\rho,\,\kappa\to 0$ limit, we expect a similar state since we can work within the regime where $\mu^{2}_{S,eff}$, $m^{2}_{S}\ll \lambda^{2}v^{2}$. However, this is not the case for finite $\kappa$, where a ``push-up" effect is expected to increase the mass of the SM-like scalar at tree level~\cite{Agashe125gev}.
% In fact, this "push-up effect" is further enhanced as $m^{2}_{S}$ becomes large. 
This effect is due to the fact that the singlet $h_{S}$ is lighter than the SM-like state and mixing between the two increases the mass of the latter. This phenomenon is evident if we write the upper $3\times 3$ mass matrix of the CP-even scalar sector in the basis $(H_{d}\cos\beta+H_{u}\sin\beta,H_{u}\cos\beta H_{d},S)\equiv (h^{0},H,h_{S})$ as in~\cite{Delgado:2012yd,Agashe125gev}, but for finite $\kappa$ and $m^{2}_{S}$ and neglect the small corrections proportional to $\rho$. In this basis, the mass matrix has the following form:
\begin{eqnarray}
\small
M^{2}_{H}=\begin{pmatrix}
m^{2}_{Z}\cos^{2}2\beta+\lambda^{2}v^{2}\sin^{2}2\beta && (m^{2}_{Z}-\lambda^{2}v^{2})\sin2\beta\cos2\beta && 2\lambda^{2}v_{S}v-2v^{2}R\\
 && m^{2}_{A}+(m^{2}_{Z}-\lambda^{2}v^{2})\sin^{2}2\beta && -2R v\cot2\beta \\
  &&    && \lambda^{2}v^{2}+m^{2}_{S}+\kappa v_{S}(4\kappa v_{S}+A_{\kappa})
\end{pmatrix}, \nonumber \\
\nonumber \\
\end{eqnarray}
where $R=\left[\frac{1}{v}\lambda(\kappa v_{S}+\frac{1}{2}A_{\lambda})\sin2\beta\right]$. Using Equation~(\ref{eq:3MinCon}) for $v_{S}$, one can see that the $(1,3)$ element of the above matrix vanishes for $\kappa ,m^{2}_{S}\to 0$. For finite $\kappa$ and $m^{2}_{S}$, the SM-like Higgs mass at tree-level is given by
\begin{equation}
m^{2}_{h^{0}}\approx m^{2}_{Z}\cos^{2}2\beta+\lambda^{2}v^{2}\sin^{2}2\beta-\frac{\left(m^{2}_{Z}-\lambda^{2}v^{2}\right)^{2}}{m^{2}_{A}}\sin^{2}2\beta\cos^{2}2\beta+\delta m^{2}_{h^{0},mix},
\end{equation}
where $\delta m^{2}_{h^{0},mix}$ is a function of $\kappa,\rho$ and $m^{2}_{S}$ and parametrizes the contribution from the $h^{0}-h_{S}$ mixing. The mixing between the SM-like Higgs, $h^{0}$, and $h_{S}$ will also have an effect on the couplings of the former to SM matter fields. In particular, it will suppress the coupling of $h^{0}$ to gauge bosons while it will enhance the coupling of $S$. Therefore, for finite $\kappa$ and $m^{2}_{S}$, hiding the light $S$ state from Higgs searches carried out by LEP becomes a strong constraint on the model's parameter space~\cite{Barate:2003sz,Abbiendi:2002qp,Schael:2006cr,Abdallah:2004wy}.

 In the limit where $\kappa, \rho\ll 1$, the mass matrix for the CP-even scalars in the basis $\left(H_{uR},H_{dR},S_{R},N_{R}\right)$ is given by the following terms:
\begin{eqnarray}
M^{2}_{11}&\approx&\lambda A_{\lambda}v_{S}\cot\beta+m^{2}_{Z}\sin^{2}\beta, \nonumber \\
M^{2}_{12}&\approx&-\lambda A_{\lambda}v_{S}-\lambda\kappa v_{S}-m^{2}_{Z}\sin\beta\cos\beta+2\lambda^{2}v^{2}\sin\beta\cos\beta-2\lambda\rho v_{N}v_{S}, \nonumber \\
M^{2}_{13}&\approx&2\lambda^{2}v_{S}v\sin\beta-\lambda A_{\lambda}v\cos\beta-2\lambda\kappa vv_{S}\cos\beta-2\lambda\rho vv_{N}\cos\beta, \nonumber \\
M^{2}_{14}&\approx&-2\lambda\rho vv_{S}\cos\beta, \nonumber \\
M^{2}_{22}&\approx&\lambda A_{\lambda}v_{S}\tan\beta+m^{2}_{Z}\cos^{2}\beta, \label{eq:cpeven} \\
M^{2}_{23}&\approx&2\lambda^{2}v_{S}v\cos\beta-\lambda A_{\lambda}v\sin\beta-2\lambda\kappa vv_{S}\sin\beta-2\lambda\rho vv_{N}\sin\beta, \nonumber \\
M^{2}_{24}&\approx&-2\lambda\rho vv_{S}\sin\beta, \nonumber \\
M^{2}_{33}&\approx&\lambda^{2}v^{2}+m^{2}_{S}+2\kappa A_{\kappa}v_{S}-\lambda\kappa v^{2}\sin2\beta+6\kappa^{2}v^{2}_{S}+\left(2A_{\rho}v_{N}+12\kappa v_{N}v_{S}\right)\rho, \nonumber \\
M^{2}_{34}&\approx&2\rho A_{\rho}v_{S}+6\kappa\rho v^{2}_{S}-\lambda\rho v^{2}\sin2\beta, \nonumber \\
M^{2}_{44}&\approx&m^{2}_{N}+4\rho^{2}v^{2}_{S}. \nonumber \label{eq:CPevenmm}
\end{eqnarray}
In the limit where $\rho\ll 0$, the mixing between the singlet $N$ and the other three scalars, $M^{2}_{i,4}$, is largely suppressed for not too large values of the tri-linear coupling $A_{\rho}$ and the $vev$ of the singlet $S$, which in this model can be adjusted through $A_{\lambda}$. The mass of $N$ will then depend mostly on the soft SUSY breaking mass parameter $m^{2}_{N}$. In our framework, we choose then to work in the following limit  $\mu^{2}_{S,eff},~m^{2}_{S} \ll \lambda^{2}v^{2}$ while keeping $m^{2}_{N}$ as a less constrained free parameter. In this limit, the masses of the CP-even scalars are given by:
\begin{eqnarray}
m^{2}_{h^{0}}&\approx& m^{2}_{Z}\cos^{2}2\beta+\lambda^{2}v^{2}\sin^{2}2\beta-\frac{\left(m^{2}_{Z}-\lambda^{2}v^{2}\right)^{2}}{m^{2}_{A}}\sin^{2}2\beta\cos^{2}2\beta+\delta m^{2}_{h^{0},mix}, \nonumber \\
m^{2}_{H}&\approx& m^{2}_{A}+\left(m^{2}_{Z}-\lambda^{2}v^{2}\right)\sin^{2}2\beta+\frac{\left(m^{2}_{Z}-\lambda^{2}v^{2}\right)^{2}}{m^{2}_{A}}\sin^{2}2\beta\cos^{2}2\beta-\frac{\lambda^{2}v^{2}A^{2}_{\lambda}}{m^{2}_{A}}\sin^{2}2\beta, \nonumber \\
m^{2}_{h_{S}}&\approx&m^{2}_{S}+\lambda^{2}v^{2}-\frac{\lambda^{2}v^{2}A^{2}_{\lambda}}{m^{2}_{A}}\cos^{2}2\beta+\delta m^{2}_{h_{S},mix}, \nonumber \\
m^{2}_{h_{N}}&\approx&m^{2}_{N}. \label{eq:CPevenMass}
\end{eqnarray}
In the above equations, we have included corrections arising from the non-decoupling of the pseudoscalar state $A$, with mass introduced in Equation~(\ref{eq:MSSMps}). The first two masses correspond to the light and heavy MSSM-like Higgs bosons. The last two correspond to the two singlet-like states. The state $h_{S}$ couples directly to the two MSSM-like states and this can be seen from the non-decoupling $1/m^{2}_{A}$ term and the term $\delta m^{2}_{h_{S},mix}$ which parametrizes the mixing between $h^{0}$ and $h_{S}$. The state $h_{N}$ is almost all singlet with its mass arising solely from the soft SUSY breaking mass parameter $m_{N}$.

The CP-odd spectrum is obtained by diagonalizing the following mass matrix in the basis $\left(H_{uI},H_{dI},S_{I},N_{I}\right)$:
\begin{eqnarray}
M^{2}_{11}&\approx&\lambda A_{\lambda}v_{S}\cot\beta, \nonumber \\
M^{2}_{12}&\approx&\lambda A_{\lambda}v_{S}+\lambda\kappa v^{2}_{S}+2\lambda\rho v_{N}v_{S}, \nonumber \\
M^{2}_{13}&\approx&\lambda A_{\lambda}v\cos\beta-2\lambda\kappa vv_{S}\cos\beta-2\lambda\rho vv_{N}\cos\beta, \nonumber \\
M^{2}_{14}&\approx&-2\lambda\rho vv_{S}\cos\beta, \nonumber \\
M^{2}_{22}&\approx&\lambda A_{\lambda}v_{S}\tan\beta,  \label{eq:cpodd} \\
M^{2}_{23}&\approx&\lambda A_{\lambda}v\sin\beta-2\lambda\kappa vv_{S}\sin\beta-2\lambda\rho vv_{N}\sin\beta, \nonumber \\
M^{2}_{24}&\approx&-2\lambda\rho vv_{S}\sin\beta, \nonumber \\
M^{2}_{33}&\approx&\lambda^{2}v^{2}+m^{2}_{S}-2\kappa A_{\kappa}v_{S}+\lambda\kappa v^{2}\sin2\beta+2\kappa^{2}v^{2}_{S}+\left(-2A_{\rho}v_{N}+4\kappa v_{N}v_{S}\right)\rho, \nonumber \\
M^{2}_{34}&\approx&-2\rho A_{\rho}v_{S}+2\kappa\rho v^{2}_{S}+\lambda\rho v^{2}\sin2\beta, \nonumber \\
M^{2}_{44}&\approx&m^{2}_{N}+4\rho^{2}v^{2}_{S}. \nonumber \label{eq:CPmm}
\end{eqnarray}
As in the CP-even sector, in addition to the MSSM-like pseudoscalar in Equation~(\ref{eq:MSSMps}), we obtain two additional singlet-like pseudoscalar states
\begin{eqnarray}
m^{2}_{A_{S}}&\approx&m^{2}_{S}+\lambda^{2}v^{2}-\frac{\lambda^{2}v^{2}A^{2}_{\lambda}}{m^{2}_{A}}, \nonumber \\
m^{2}_{A_{N}}&\approx&m^{2}_{N}.\label{eq:CPoddMass}
\end{eqnarray}
In order to keep the above equations as clear and simple as possible, we have not incorporated corrections proportional to $\rho$. However, the calculation of the masses is done exactly in our numerical routines.

As mentioned earlier in this section, in order to generate an spectrum similar to the one studied in~\cite{Delgado:2012yd}, it is important to work in the limit where $m^{2}_{S}\ll \lambda^{2}v^{2}$. This condition is somewhat unnatural since there exist contributions to the one-loop renormalization group equation for $m^{2}_{S}$ that are proportional to $A^{2}_{\lambda}$~\cite{Ellwanger:2008py}, that in our framework is large in order to decouple the MSSM-like pseudoscalar, with mass given in Equation~(\ref{eq:MSSMps}), from the spectrum. One may alleviate this by embedding the model into a SUSY breaking mediation mechanism where the scale of SUSY breaking is not very high. 
%%%%%%%%%%%%%%%%%%%%%%%%%%%%%%%%%%%%%%%%%%%%%%%%%%%%%%%%%%%%%%%%%%%%%%%%%%%%%%%%%%%%%%%%%%%%%%%%

\subsection{Neutralino Sector}

The neutral gauginos of this model $(\tilde{B},\,\tilde{W}^0)$ mix with the two neutral higssinos and the two singlinos to form the neutralino mass eigenstates due to the electroweak symmetry breaking and the Yukawa couplings. Using the basis 
\begin{equation}
\psi^0=(\tilde{B},\tilde{W}^0,\tilde{H}^0_d,\tilde{H}^0_u, \tilde{S},\tilde{N}) \nonumber , 
\end{equation}
 the mass terms in the Lagrangian for the neutralino sector are given by 
  \begin{equation}
 -\frac{1}{2} \psi^{0\,T} M_{\tilde{N}} \psi^0 + {\rm c.c.}, 
 \end{equation} where 
\begin{equation}
M_{\tilde{N}}=\left(
\begin{array}{cccccc}
 M_1 & 0 & -\frac{g' m_W \cos (\beta )}{g} & \frac{g' m_W \sin (\beta )}{g} & 0 & 0 \\
 0 & M_2 & m_W \cos (\beta ) & -m_W \sin (\beta ) & 0 & 0 \\
 -\frac{g' m_W \cos (\beta )}{g} & m_W \cos (\beta ) & 0 & -\lambda  v_s & -\lambda  v_s \sin (\beta ) & 0 \\
 \frac{g' m_W \sin (\beta )}{g} & -m_W \sin (\beta ) & -\lambda  v_s & 0 & -\lambda  v_s \cos (\beta ) & 0 \\
 0 & 0 & -\lambda  v_s \sin (\beta ) & \lambda  v_s \cos (\beta ) & 2 \rho  v_n+2 \kappa  v_s & 2 \rho  v_s \\
 0 & 0 & 0 & 0 & 2 \rho  v_s & 0 \\ 
\end{array}
\right) \label{M_neut}.
\end{equation}
The corresponding mass eigenstates are given by  
\begin{equation}
\chi^0_i = N_{i\,j} \psi^0_j, \label{neut_mass} 
\end{equation}
where the unitary mixing matrix, $N_{i\,j}$, diagonalizes Equation~(\ref{M_neut}),
\begin{equation}
N^*\, M_{\tilde{N}}\, N^{-1} = {\rm diag}(m_{\chi^0_1},m_{\chi^0_2},m_{\chi^0_3},m_{\chi^0_4},m_{\chi^0_5},m_{\chi^0_6}),\label{eq:NeutDiag}
\end{equation} and where the eigenmasses have been labeled in ascending order. 

The leading contributions to the masses of the two lightest neutralinos, $\chi^0_{1,2}$, are given by
\begin{eqnarray}
m_{\chi^0_1} &\approx& \left( (\kappa v_S+ \rho v_N)^2 +4 \rho^2 v_S^2\right)^{1/2} - |\kappa v_S + \rho v_N |, \\
m_{\chi^0_2} &\approx& \left( (\kappa v_S+ \rho v_N)^2 +4 \rho^2 v_S^2\right)^{1/2} + |\kappa v_S + \rho v_N |.
\end{eqnarray}

Within our framework, $\chi^0_1$ is mostly singlino and couples weakly to the other particles in the spectrum. However, as it will be shown in Section \ref{sec:Constraints}, it could have a small but significant bino and higgsino components. This will play an important role in the cosmological evolution of the energy density of this stable particle, since it makes the self-annihilation effective enough to avoid overabundance of the relics. On the other hand, the mass of the next-to-lightest neutralino within the parameter space considered in Section \ref{sec:Constraints}, will be at least twice as massive as the lightest supersymmetric particle (LSP); and co-annihilations between $\chi^0_1$ and the heavier neutralino will not relevant for the calculation of its relic abundance. Instead, the relic density will be determined by the annihilation cross-section of the LSP, as explained in Section \ref{sec:DarkMatter}. The couplings of the lightest neutralino to the CP-odd and CP-even Higss scalars, that will be used in the relic abundance calculation, are given by
\begin{eqnarray}
C_S^{\chi\chi\, h_j}&= &\frac{2}{\sqrt{2}}\left[-\lambda {\cal N}_{56}{\cal N}_{46} R^{H}_{j2}-\lambda {\cal N}_{56}{\cal N}_{36}R^{H}_{j1}+(\kappa {\cal N}_{56}{\cal N}_{56}+2\rho {\cal N}_{66}{\cal N}_{56}-\lambda {\cal N}_{36}{\cal N}_{46})R^{H}_{j3}\right.\nonumber \\
&+&\left.\rho {\cal N}_{56}{\cal N}_{56}R^{H}_{j4}+\left(\frac{g}{\sqrt{2}}{\cal N}_{46}{\cal N}_{26}-\frac{g'}{\sqrt{2}}{\cal N}_{46}{\cal N}_{16}\right)R^{H}_{j1}\right. \nonumber \\
&+&\left.\left(\frac{g'}{\sqrt{2}}{\cal N}_{36}{\cal N}_{16}-\frac{g}{\sqrt{2}}{\cal N}_{36}{\cal N}_{26}\right)R^{H}_{j2}\right], \nonumber \\ 
& & \nonumber \\ 
C_P^{\chi\chi\, A_i} &= &\frac{2}{\sqrt{2}}\left[-\lambda {\cal N}_{56}{\cal N}_{46} R^{A}_{i2}-\lambda {\cal N}_{56}{\cal N}_{36}R^{A}_{i1}+(\kappa {\cal N}_{56}{\cal N}_{56}+2\rho {\cal N}_{66}{\cal N}_{56}-\lambda {\cal N}_{36}{\cal N}_{46})R^{A}_{i3}\right.\nonumber \\
&+&\left.\rho {\cal N}_{56}{\cal N}_{56}R^{A}_{i4}+\left(\frac{g}{\sqrt{2}}{\cal N}_{46}{\cal N}_{26}-\frac{g'}{\sqrt{2}}{\cal N}_{46}{\cal N}_{16}\right)R^{A}_{i1}\right. \nonumber \\
&+&\left.\left(\frac{g'}{\sqrt{2}}{\cal N}_{36}{\cal N}_{16}-\frac{g}{\sqrt{2}}{\cal N}_{36}{\cal N}_{26}\right)R^{A}_{i2}\right], \label{LSPCoupl}
\end{eqnarray} where $i=1,2,3$, $j=1,2,3,4$, ${\cal N} \equiv N^{-1}$ and $R^A$ ,$R^H$ diagonalize the CP-odd and CP-even mass matrices defined in Equations~(\ref{eq:CPevenmm}) and~(\ref{eq:CPmm}).

%%%%%%%%%%%%%%%%%%%%%%%%%%%%%%%%%%%%%%%%%%%%%%%%%%%%%%%%%%%%%%%%%
%
%
%
%
%
%%%%%%%%%%%%%%%%%%%%%%%%%%%%%%%%%%%%%%%%%%%%%%%%%%%%%%%%%%%%%%%%%
\section{Constraints and parameter scan}\label{sec:Constraints}

\subsection{LEP Constraints}

One important constraint on the parameter space is due to the LEP bound on the chargino mass, $m_{\chi^{+}}>104$ GeV. This bound translates into a bound on $\mu_{eff}$ given by $|\mu_{eff}|>104$ GeV. Using Equation~(\ref{eq:3MinCon}), this can be re-casted into a bound on $A_{\lambda}$ given by $A_{\lambda}>208\frac{\left(1+m^{2}_{S}/\lambda^{2}v^{2}\right)}{\sin2\beta}$ GeV. For $\tan\beta=2$ and $m^{2}_{S}\ll\lambda^{2}v^{2}$, $A_{\lambda}$ is bounded from below by $260$ GeV. However, constraints on the singlet-like scalar fields yield a finite value for $m^{2}_{S}$ and the lower bound for $A_{\lambda}$ lies slightly above $260$ GeV. 

Constraints on light scalars also limit the parameter space of this model. In particular, searches by LEP~\cite{Barate:2003sz,Abbiendi:2002qp,Schael:2006cr,Abdallah:2004wy} place strong upper bounds on the two main scalar production mechanisms: $e^{+}e^{-}\to HZ$ and  $e^{+}e^{-}\to HA$, where $H$ and $A$ denote any of the CP-even or -odd scalars respectively. In the $HZ$ channel these constraints assume that each scalar decays to $b\bar{b}$ or $\tau^{+}\tau^{-}$ with a branching fraction equal to one. In general, these bounds will soften since the scalars in our framework can also decay to lighter scalars with a significant branching fraction.

The possibility of Higgs cascade decays has also been searched for at LEP~\cite{Schael:2006cr,Abdallah:2004wy}. They place strong bounds on two channels: (1) Associated Higgs production with a $Z$, $e^{+}e^{-}\to ZH_{i}$, $H_{i}\to A_{j}A_{j}$ and (2) Scalar-pseudoscalar pair production, $e^{+}e^{-}\to H_{i}A_{j}$. In (1), the analysis assumes a $BR(H_{j},A_{j}\to b\bar{b})=1$ and $BR(H_{i}\to H_{j}H_{j},A_{j}A_{j})=1$. In (2), five different final states were analyzed:
\begin{eqnarray}
e^{+}e^{-}&\to& H_{i}A_{j}\to 4b, \nonumber \\
e^{+}e^{-}&\to& H_{i}A_{j}\to 4\tau, \nonumber \\
e^{+}e^{-}&\to&H_{i}A_{j}\to A_{k}A_{k}A_{j}\to 6b, \nonumber \\
e^{+}e^{-}&\to&H_{i}A_{j}\to A_{k}A_{k}A_{j}\to 6\tau, \nonumber \\
e^{+}e^{-}&\to& H_{i}A_{j}\to 2b,2\tau. \nonumber \\
\end{eqnarray}
In our analysis, we calculate the normalized cross section for scalar-pseudoscalar pair production which is given by 
\begin{equation}
\sigma_{H_{i}A_{i}}=\bar{\lambda}\sigma^{SM}_{HZ},
\end{equation}
where $\bar{\lambda}$ is a kinematic factor given by
\begin{eqnarray}
\bar{\lambda}=\lambda^{3/2}_{A_{i}H_{j}}\left[\lambda^{1/2}_{ZH_{j}}(12m^{2}_{Z}/s+\lambda_{ZH_{i}})\right], \nonumber \\
\lambda_{ij}=\left[1-(m_{i}+m_{j})^{2}/s\right]\left[1-(m_{i}-m_{j})^{2}/s\right],
\end{eqnarray}
and $s$ is the center of mass energy squared. We multiply the normalized cross-section, $\sigma_{H_{i}A_{i}}/\sigma^{SM}_{HZ}$, by the appropriate branching fractions in the decay chain. Furthermore, we implement the constraint found in the channel $e^{+}e^{-}\to ZH$ that is independent of the H decay mode~\cite{Abbiendi:2002qp}.

%%%%%%%%%%%%%%%%%%%%%%%%%%%%%%%%%%%%%%%%%%%%%%%%%%%%%%%%%%%%%%%%%%%%%%%%%%%%%%%%%%%%%%%%%%%%%%%%%%%%%%%%%%%%%%%%
\subsection{Meson Decays}
A pseudoscalar, with a mass in the range between $1$ and $40$ GeV, has a coupling to fermions that is highly constrained by meson decays and collider data. The couplings can be extracted from the following Lagrangian:
\begin{equation}
{\cal L}\supset -i\frac{g}{2m_{W}}A_{i}\left(C_{A_{i}uu}m_{u}\bar{u}\gamma^{5}u+C_{A_{i}dd}m_{d}\bar{d}\gamma^{5}d+C_{A_{i}ll}m_{l}\bar{l}\gamma^{5}l\right),
\end{equation}
where
\begin{eqnarray}
C_{A_{i}uu}=R^{A}_{1i}\cot\beta, \nonumber \\
C_{A_{i}dd}=C_{A_{i}ll}=R^{A}_{1i}\tan\beta,
\end{eqnarray}
denote the couplings of the pseudoscalar mass eigenstates, $A_{i}$, to $up$-type and $down$-type quarks respectively and $R^{A}_{ij}$ is the unitary matrix that diagonalizes the CP-odd mass matrix introduced in Equation~(\ref{eq:CPmm}). For masses below the upsilon threshold of $\sim 9.46$ GeV, an analysis by~\cite{Dermisek:2010mg} found that $\Upsilon\to\gamma A_{i}$ imposes that $C_{A_{i}dd}<0.5$ for $\tan\beta\sim 1$. Above this mass threshold the same analysis found the strongest constraint on the pseudoscalar mass coming from the process $e^{+}e^{-}\to b\bar{b} A_{i}\to b\bar{b} b\bar{b}$ measured by DELPHI~\cite{Abdallah:2004wy}, setting the following limit, $C_{A_{i}dd}< O\left(10\right)$. Additional constraints on light pseudoscalars arise form rare $B$ and $K$ decays such as: $B\to K+~\text{invisible},~K\to \pi+~\text{invisible},~B\to K e^{+}e^{-},~K\to \pi e^{+}e^{-}$ and $K\to \pi^{+}+X$ as well as the muon $g-2$ . The analysis in~\cite{Andreas:2010ms} on an NMSSM light pseudoscalar concludes that pseudoscalar masses of $m_{A_{i}}<2m_{\mu}$ are excluded unless the coupling $C_{A_{i}dd}$ lies below $10^{-4}$. Within our framework, the parameter space consistent with bounds on light scalars and supersymmetric particles yields pseudoscalar masses above the $2m_{\mu}$ threshold.

\subsection{LHC Constraints}

The discovery of a SM-like Higgs boson with mass around $126$ GeV provides a new set of constraints that must be addressed in order for the known production cross sections and decay rates to be in agreement with those measured at the LHC~\cite{Aad:2012tfa,Chatrchyan:2013lba}. The authors in~\cite{Barger:2012hv} have proposed a method of calculating the total width of a SM-like Higgs boson using data from the LHC and the Tevatron as well as the properties of the SM-like Higgs boson as a benchmark. Furthermore, they provide a method for estimating the branching fraction of the SM-like Higgs boson to dark matter. In addition, the authors in~\cite{Belanger:2013xza} have carried out a global fit to the data and found a total width of a Higgs relative to the SM prediction given by $\Gamma_{tot}/\Gamma^{SM}_{tot}\in[0.5,2]$ and an invisible branching fraction of roughly $38\%$ at $95\%$ CL. These results were obtained by varying the Higgs couplings to SM particles independently of each other. More conservative results were obtained by setting the couplings of the Higgs to SM particles to their SM values. They find a $\Gamma_{tot}/\Gamma^{SM}_{tot}\in [1,1.25]$ and $Br(h^{0}\to inv)\le19\%$ at $95\%$ CL. In our analysis we calculate the total width of the SM-like Higgs boson, since this gets contributions from light singlet-like scalars and pseudoscalars as well as the light singlet-like neutralinos, and look for deviation from the SM value of $\Gamma^{SM}_{tot}=4.1$ MeV~\cite{Dobrescu:2012td}. We require that $0.5\le\Gamma_{tot}/\Gamma^{SM}_{tot}\le2$ and a Higgs invisible branching fraction of $Br(h^{0}\to inv)\lesssim40\%$.

\subsection{$\Gamma^{inv}_{Z}$ and Neutralino sector}
The neutralino sector of this model contains two states with a large singlet component, however, the next-to-lightest neutralino may also have a significant amount of Higgsino component. If this neutralino is lighter than $m_{Z}/2$, $Z$ decays to a pair of next-to-lightest neutralinos could violate bounds on the invisible decay width of the $Z$. The decay of the $Z$ into a pair of neutralinos is given by:
\begin{equation}
\Gamma_{Z\to\chi_{n}\chi_{n}}=\frac{(g'^{2}+g^{2})}{4\pi}\frac{(|N_{n,3}|^{2}-|N_{n,4}|^{2})^{2}}{24m^{2}_{Z}}\left(m^{2}_{Z}-2m^{2}_{\chi_{n}}\right)^{3/2},
\end{equation}
where $N_{n,3}$ and $N_{n,4}$ are the down- and up-type Higgsino components of the n$^{th}$ neutralino mass eigenstate respectively as described in Section~\ref{sec:TheModel}. However, the bound on the invisible $Z$ decay width, $\Delta\Gamma^{inv}_{Z}<2.3$ MeV~\cite{ALEPH:2005ab}, sets more stringent constraints on the next-to-lightest neutralino since it has a larger higgsino component. Furthermore, we find that the next-to-lightest neutralino has a mass below $90$ GeV and thus the production process $e^{+}e^{-}\to\chi^{0}_{1}\chi^{0}_{2}$ was kinematically accessible at LEP 2. The strongest bound was found by the OPAL collaboration~\cite{Abbiendi:2003sc}. Since we are considering a lightest neutralino with a mass below $20$ GeV, the cross section for the process $e^{+}e^{-}\to\chi^{0}_{1}\chi^{0}_{2}$ is bounded from above by $0.05$ pb. To calculate the cross section we follow the analysis in~\cite{Das:2010ww} where
\begin{equation}
\sigma_{Z}(e^{+}e^{-}\to\chi^{0}_{1}\chi^{0}_{2})~\text{[pb]}\approx4.9\times 10^{4}\frac{(s-m^{2}_{\chi^{0}_{2}})^{2}}{s(s-m^{2}_{Z})^{2}}\left(1+\frac{m^{2}_{\chi^{0}_{2}}}{2s}\right)\left(N_{13}N_{23}-N_{14}N_{24}\right)^{2},
\end{equation}
$s=209.^{2}$ GeV$^{2}$ is the center of mass energy at LEP 2 and $N_{ij}$ is the matrix that diagonalizes the neutralino weak eigenstates introduced in Equation~(\ref{eq:NeutDiag}).

%%%%%%%%%%%%%%%%%%%%%%%%%%%%%%%%%%%%%%%%%%%%%%%%%%%%%%%%%%%%%%%%%%%%%%%%%%%%%%%%%%%%%%%%%%%%%%%%%%%%%%
\subsection{Parameter scan considerations}

We analyze the parameter space of this model necessary to generate a SM-like Higgs boson with a mass of $126$ GeV and light singlet-like states that are consistent with Higgs searches carried out by LEP~\cite{Barate:2003sz,Abbiendi:2002qp,Schael:2006cr,Abdallah:2004wy}. The SM-like Higgs mass is given by:
\begin{equation}
m^{2}_{h^{0}}\approx m^{2}_{Z}\cos^{2}2\beta+\lambda^{2}v^{2}\sin^{2}2\beta-\frac{\left(m^{2}_{Z}-\lambda^{2}v^{2}\right)^{2}}{m^{2}_{A}}\sin^{2}2\beta\cos^{2}2\beta+\delta m^{2}_{h^{0},mix}+\delta m^{2}_{h^{0},loop},
\end{equation}
where $\delta m^{2}_{h^{0},loop}$ parametrizes the leading radiative corrections to the SM-like Higgs mass from third generation of quarks/squarks. This correction is given by
\begin{equation}
\delta m^{2}_{h^{0},loop}=\frac{3\bar{m_{t}}^{2}}{2\pi^{2}v^{2}}\left[\log\frac{M_{\tilde{t}}}{m_{t}}+\frac{X_{t}}{4}+\frac{\log\frac{M_{\tilde{t}}}{m_{t}}}{32\pi^{2}}(3m^{2}_{t}/v^{2}-16g^{2}_{s})\left((X_{t}+2\log\frac{M_{\tilde{t}}}{m_{t}}\right)\right],
\end{equation}
where $\bar{m}_{t}=m_{t}/(1+4\alpha_{s}/3\pi)$, $m_{t}$ is the pole mass of the top quark, $g_{s}$ is the strong coupling constant, $M_{\tilde{t}}$ is the geometric mean of the two top squark mass eigenvalues and $X_{t}$ parametrizes the mixing between top squarks:
\begin{equation}
X_{t}=\frac{2(A_{t}-\mu/\tan\beta)^{2}}{M^{2}_{\tilde{t}}}\left[1-\frac{(A_{t}-\mu/\tan\beta)^{2}}{12M^{2}_{\tilde{t}}}\right].
\end{equation}
In order to maximize this value at tree level, we consider large values of $\lambda$. However, we insist that $\lambda$ remains perturbative at all scales up to the grand unification scale $M_{GUT}=2\times10^{16}$ GeV. This places an upper bound on $\lambda$ which peaks for values of $\tan\beta$ between 2 and 3 as in the models described in~\cite{Ellwanger:2008py,Delgado:2010uj,Delgado:2012yd}. Our analysis is carried out with $\tan\beta=2$.

\begin{table}[ht]\centering
\begin{tabular}{|c|c|c| }
\hline
\hline
       & Description & Range \\
\hline
$A_{t}$ & SUSY-breaking top trilinear coupling & $[0,1000]$ GeV \\
\hline
$m^{2}_{\tilde{t}_{L}}$ & Soft mass for left handed stop & $[650^{2},1000^{2}]$ GeV$^{2}$ \\
\hline
$M_{2}$ & Wino mass & $[250,2500]$ GeV \\
\hline
$\lambda$ & $\hat{S}-\hat{H}_{u}-\hat{H}_{d}$ trilinear coupling & $[0.5,0.63]$ \\
\hline
$\kappa$ & Singlet self coupling & $[-0.1,0.1]$ \\
\hline
$\rho$ & $\hat{S}-\hat{N}$ superpotential coupling & $[-0.05,0.05]$ \\
\hline
$A_{\lambda}$ & SUSY-breaking $S-H_{u}-H_{d}$ trilinear coupling & $[0,1000]$ GeV \\
\hline
$A_{\kappa}$ & SUSY-breaking single trilinear coupling & $[0,500]$ GeV  \\
\hline
$A_{\rho}$ & SUSY-breaking $N-S^{2}$ trilinear coupling & $[0,500]$ GeV \\
\hline
$m^{2}_{S}$ & SUSY-breaking mass term for $S$ & $[0,1000]$ GeV$^{2}$ \\
\hline
$M^{2}_{N}$ & SUSY-breaking mass term for $N$ & $[0,1000]$ GeV$^{2}$ \\
\hline
\end{tabular}

\caption{ Model parameters and their ranges used in the numerical routine.} \label{tab:TabParam}
\end{table}
Our calculations of the Higgs masses are done using a full one-loop effective potential. Furthermore, in order to maximize the SM-like Higgs mass we use a large MSSM-like pseudoscalar mass, $m_{A}$. This has the effect of decoupling one of the Higgs doublets from the Higgs sector. In the analysis, we use the four minimization conditions introduced in Equations~(\ref{eq:1MinCon})-(\ref{eq:4MinCon}) and solve for $m^{2}_{H_{u}},m^{2}_{H_{d}},v_{S}$ and $v_{N}$. The remaining parameters of the model are varied as in Table~\ref{tab:TabParam}. In the scan we fix the mass of the Bino at half the Wino mass, and we set the gluino mass at $3.0$ TeV. Based on the constraints introduced in the previous sections we focus on a subset of the parameter scan introduced in Table~\ref{tab:TabParam}. We choose a benchmark point that does not introduce a large amount of fine tuning in the stop sector, generates light scalar/pseudoscalar states, heavy Higgsino-like neutralinos as well  as decouples the heavy MSSM-like scalar and pseudoscalar states. The parameters chosen for this benchmark scan are given in Table~\ref{tab:TabParam2}. The remaining three parameters, $\kappa$, $\rho$, and $M_{2}$ are scanned keeping in mind the following considerations:
\begin{table}[ht]\centering
\begin{tabular}{|c|c|c| }
\hline
\hline
       & Description & Value \\
\hline
$A_{t}$ & SUSY-breaking top trilinear coupling & $700$ GeV \\
\hline
$m_{\tilde{t}_{L,R}}$ & Soft mass for left- and right-handed stops & $700$ GeV$$ \\
\hline
$\lambda$ & $\hat{S}-\hat{H}_{u}-\hat{H}_{d}$ trilinear coupling & $0.57$ \\
\hline
$A_{\lambda}$ & SUSY-breaking $S-H_{u}-H_{d}$ trilinear coupling & $900$ GeV \\
\hline
$A_{\kappa}$ & SUSY-breaking single trilinear coupling & $100$ GeV  \\
\hline
$A_{\rho}$ & SUSY-breaking $N-S^{2}$ trilinear coupling & $200$ GeV \\
\hline
$m^{2}_{S}$ & SUSY-breaking mass term for $S$ & $2000$ GeV$^{2}$ \\
\hline
$M^{2}_{N}$ & SUSY-breaking mass term for $N$ & $5000$ GeV$^{2}$ \\
\hline
\end{tabular}

\caption{ Model parameters and their values used in the sub-scan.} \label{tab:TabParam2}
\end{table}
\begin{itemize}
\item
$\kappa$ is scanned in order to minimize the invisible branching fraction contribution to the total width of the SM-like Higgs boson. The coupling of neutralinos to the SM-like Higgs boson is given by:
\begin{eqnarray}
g_{h^{0},\chi_{i},\chi_{j}}&\approx&\frac{1}{\sqrt{2}}\left[-\lambda {\cal N}_{5i}{\cal N}_{4j}R^{H}_{22}-\lambda {\cal N}_{5i}{\cal N}_{3j}R^{H}_{21}+(\kappa {\cal N}_{5i}{\cal N}_{5j}+2\rho {\cal N}_{6i}{\cal N}_{5j}-\lambda {\cal N}_{3i}{\cal N}_{4j})R^{H}_{23}\right.\nonumber \\
&+&\left.\rho {\cal N}_{5i}{\cal N}_{5j}R^{H}_{24}+\left(\frac{g}{\sqrt{2}}{\cal N}_{4i}{\cal N}_{2j}-\frac{g'}{\sqrt{2}}{\cal N}_{4i}{\cal N}_{1j}\right)R^{H}_{21}\right. \nonumber \\
&+&\left.\left(\frac{g'}{\sqrt{2}}{\cal N}_{3i}{\cal N}_{1j}-\frac{g}{\sqrt{2}}{\cal N}_{3i}{\cal N}_{2j}\right)R^{H}_{22}\right] \nonumber \\
&+&\frac{1}{\sqrt{2}}\left[-\lambda {\cal N}_{5j}{\cal N}_{4i}S^{H}_{22}-\lambda {\cal N}_{5j}{\cal N}_{3i}R^{H}_{21}+(\kappa {\cal N}_{5j}{\cal N}_{5i}+2\rho {\cal N}_{6j}{\cal N}_{5i}-\lambda {\cal N}_{3j}{\cal N}_{4i})R^{H}_{23}\right.\nonumber \\
&+&\left.\rho {\cal N}_{5j}{\cal N}_{5i}R^{H}_{24}+\left(\frac{g}{\sqrt{2}}{\cal N}_{4j}{\cal N}_{2i}-\frac{g'}{\sqrt{2}}{\cal N}_{4j}{\cal N}_{1i}\right)R^{H}_{21}\right. \nonumber \\
&+&\left.\left(\frac{g'}{\sqrt{2}}{\cal N}_{3j}{\cal N}_{1i}-\frac{g}{\sqrt{2}}{\cal N}_{3j}{\cal N}_{2i}\right)R^{H}_{22}\right],
\end{eqnarray}
where $R^{H}_{ij}$ is a unitary matrix that diagonalizes the CP-even mass matrix in Equation~(\ref{eq:CPevenmm}) and ${\cal N}$ is the inverse of $N$ which was introduced in Equation~(\ref{neut_mass}) and diagonalizes the neutralino sector. Additionally, $\kappa$ sets the mass of the next-to-lightest neutralino, which in our model sits well above the lightest neutralino mass.
\item
$\rho$ is scanned in order to generate a lightest neutralino with a mass below $\sim 15$ GeV. We also use a small value of $\rho$ such that the lightest scalar/pseudoscalar in the spectrum have very little mixing with the MSSM-like scalar/pseudoscalar states.
\end{itemize}

\begin{figure}[h]\centering
\bigskip
\subfigure{\includegraphics[width=2.9in]{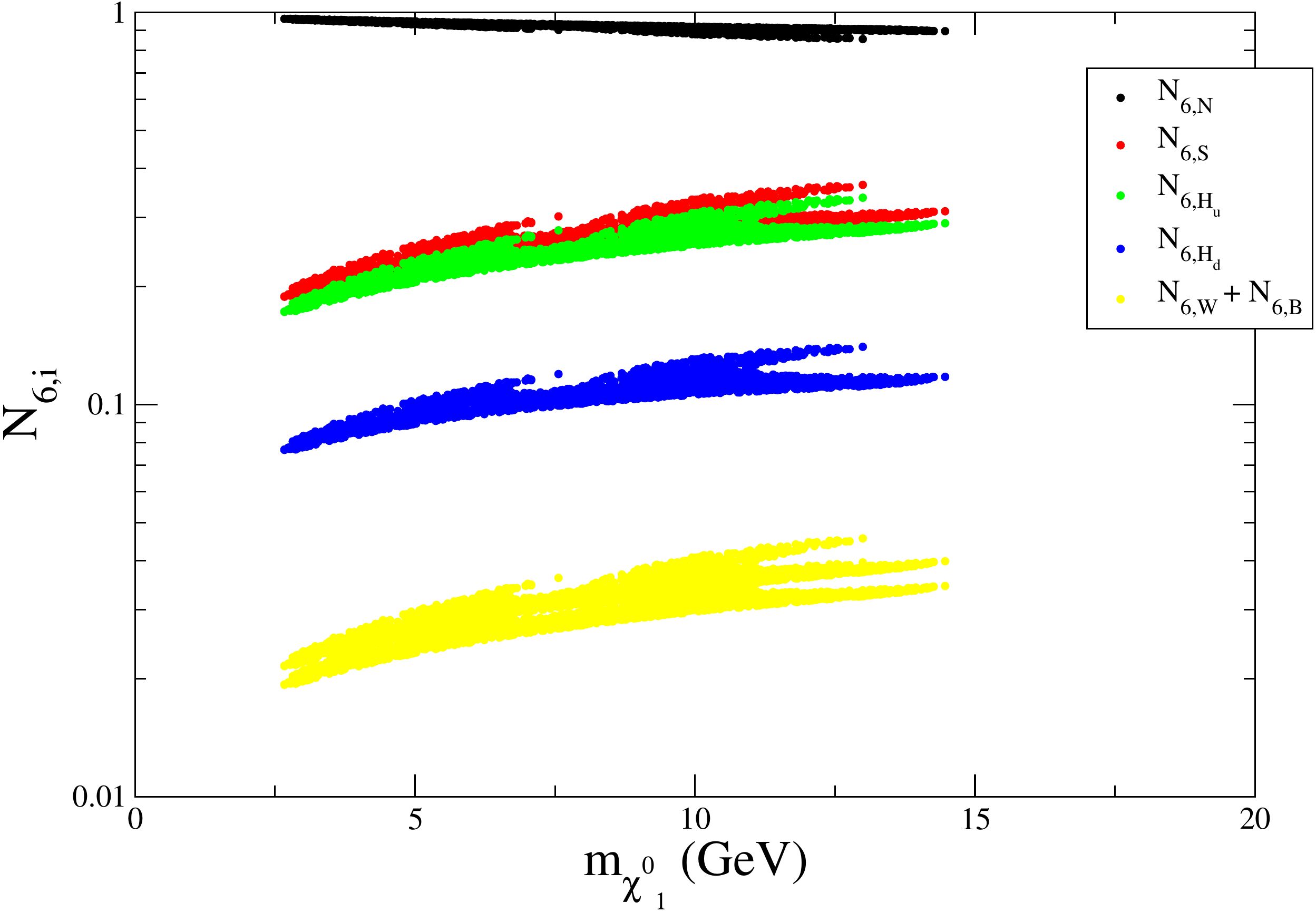}}~
\hskip 0.2in
\subfigure{\includegraphics[width=2.9in]{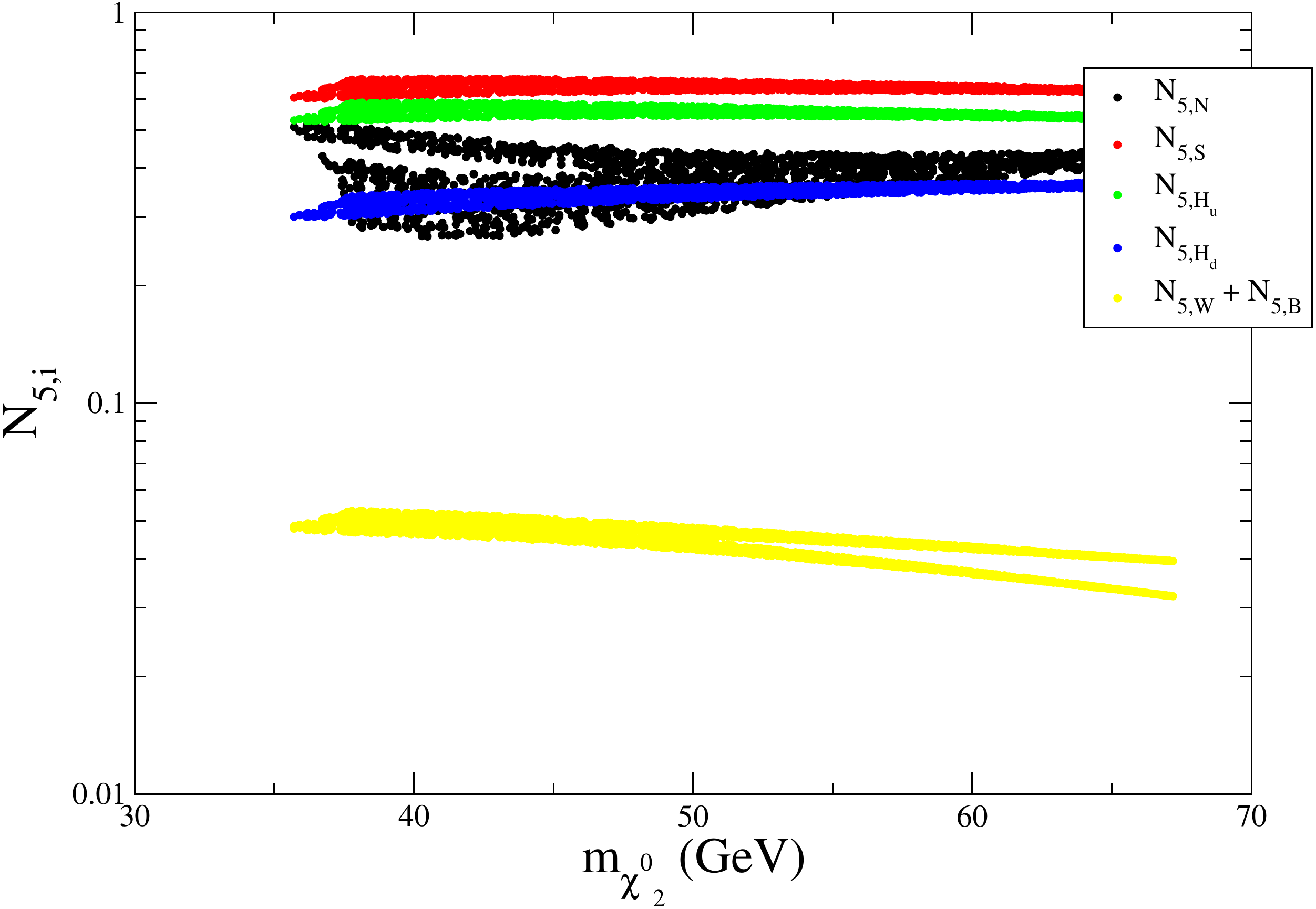}}
\caption{\small The different components of the lightest neutralino on the right and the next-to-lightest neutralino for a Wino mass parameter $M_{2}=500$ GeV.} \label{fig:Fig1}
%\big-skip
\end{figure}
The value of $\kappa$ is scanned between $-0.1$ and $-0.01$ and $\rho$ between $0.01$ and $0.05$. We run our numerical routines considering two values of the Wino mass, $[500,1500]$ GeV. In Figure~\ref{fig:Fig1} we show the different components of the lightest neutralino (left figure) and the next-to-lightest neutralino (right figure) for $M_{2}=500$ GeV. Both figures are consistent with a Higgs mass of roughly $126$ GeV, the invisible $Z$ width, and LEP bounds on charginos. The contribution to the invisible decay width of the SM-like Higgs will arise mainly from the $h^{0}\to\chi^{0}_{2}\chi^{0}_{2}$ and $h^{0}\to\chi^{0}_{1}\chi^{0}_{2}$ decay channels. This is due to the fact that the next-to-lightest neutralino has a large amount of mixing with the Higgsinos and a fine cancellation between the parameters in the model, $\lambda, \kappa$ and $\rho$, is needed. Furthermore, if we compare Figure~\ref{fig:Fig1} which corresponds to $M_{2}=500$ GeV with Figure~\ref{fig:Fig2} which corresponds to $M_{2}=1.5$ TeV, the next-to-lightest neutralino in the former has a larger component along the Wino and Bino directions. Therefore, one will expect that for $M_{2}=500$ GeV, the values of $\kappa$ and $\rho$ are more fine tuned for the model to satisfy the constraints from the invisible width of the Higgs as well all other LEP constraints introduced in the previous sections. In fact, this can be seen in Figures~\ref{fig:pseudo1500} and~\ref{fig:DM1500} which correspond to contours of the lightest pseudoscalar and lightest neutralino masses in the $\rho-\kappa$ plane. Both figures were obtained with $M_{2}=1.5$ TeV. Within the plot on the left, only the SM-like Higgs mass constraint, the invisible $Z$ width, and the chargino mass bound were taken into consideration. The plot on the right was obtained after applying the entire set of constraints. It is evident from the figures that the range of $\kappa$ becomes narrow as $\rho$ changes. This is due to the fact that some cancellations have to happen between $\lambda,\kappa$ and $\rho$ in order for $Br(h^{0}\to inv)\lesssim 40\%$. However, we still manage to get a large enough range of $\chi^{0}_{1}$ masses for a wide enough range of $\kappa$ and $\rho$ parameter values. This is also true for the mass of the lightest pseudoscalar, and as we will show in the next section, the annihilation $\chi^{0}_{1}\chi^{0}_{1}\to A_{N}\to \bar{l}l,\bar{q}q$ can be efficient enough to generate the right density of dark matter. The situation is a bit more constrained for $M_{2}=500$ GeV. In this case, the next-to-lightest neutralino has a larger wino and bino component and a finer cancellation among parameters is necessary to satisfy the constraint on the invisible decay width of $h^{0}$. Within this benchmark scenario, after all constraints have been applied, the lightest neutralino has a mass of $8$ GeV and the allowed values for $\kappa$ and $\rho$ are $-0.05$ and $0.023$ respectively. The corresponding value of the lightest pseudoscalar mass is $30$ GeV.
\begin{figure}[ht]\centering
\bigskip
\vspace{0.5cm}
\subfigure{\includegraphics[width=2.9in]{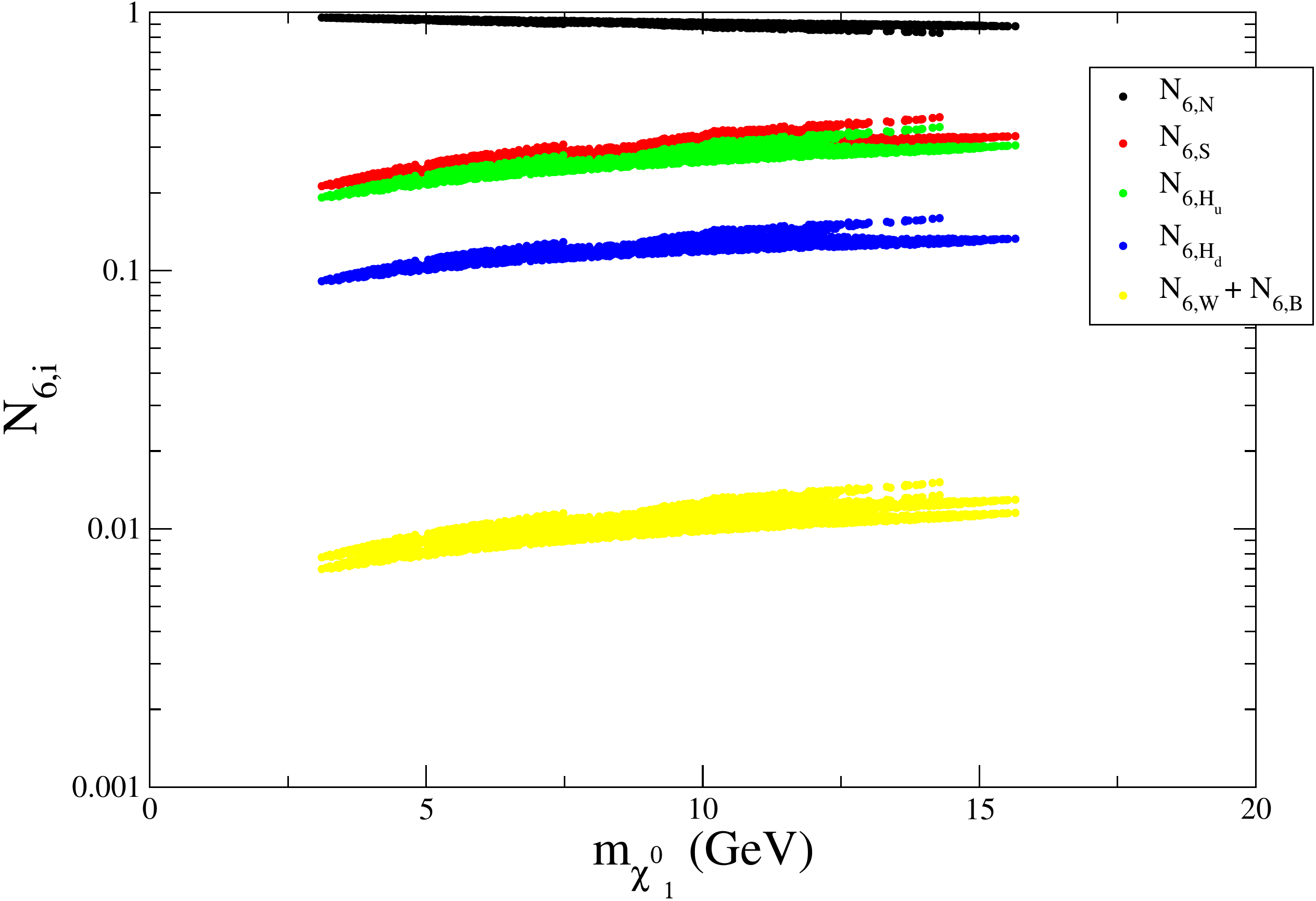}}~
\hskip 0.2in
\subfigure{\includegraphics[width=2.9in]{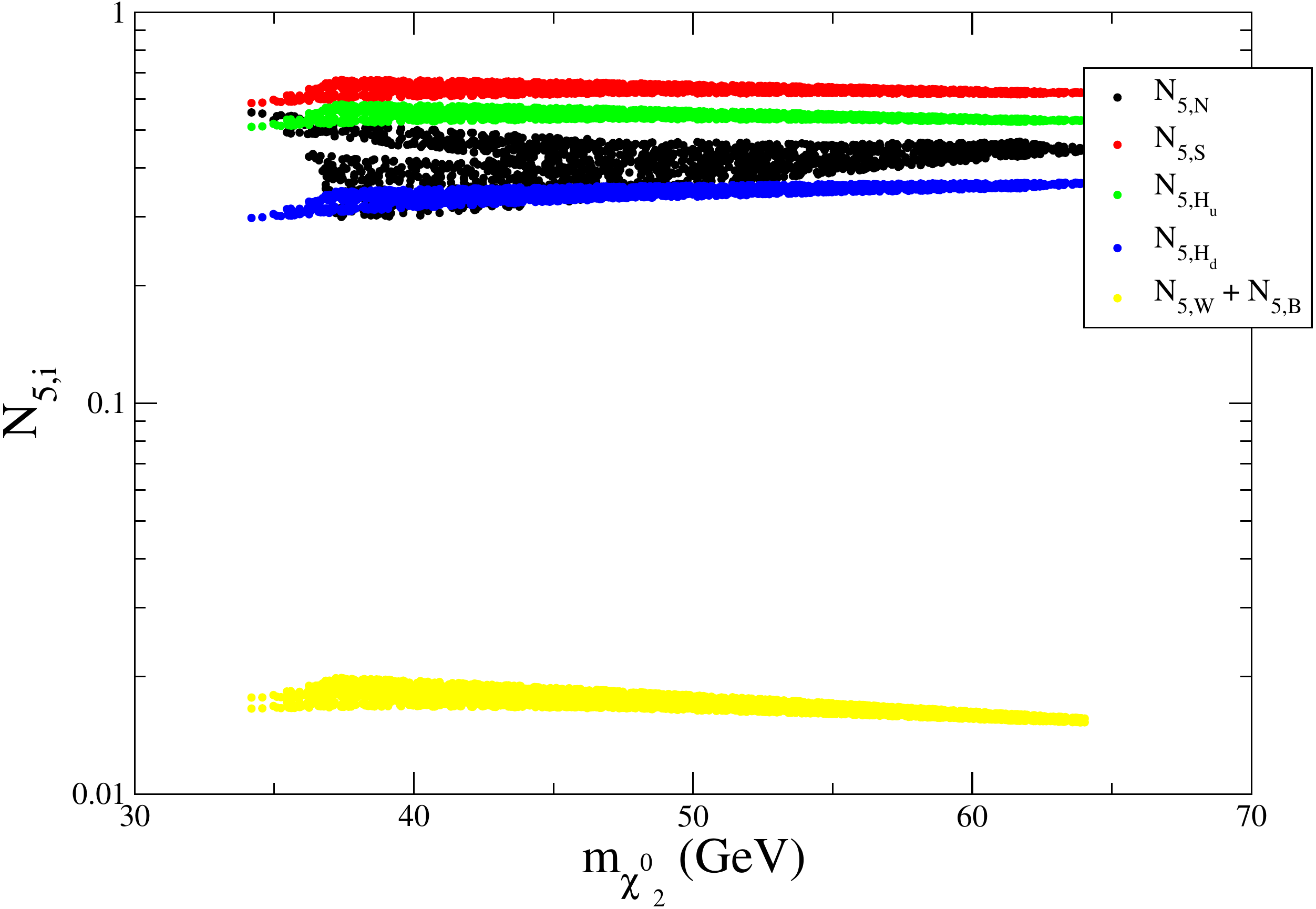}}
\caption{\small The different components of the lightest neutralino on the right and the next-to-lightest neutralino for a Wino mass parameter $M_{2}=1.5$ TeV.} \label{fig:Fig2}
\bigskip
\end{figure}

\begin{figure}[ht]\centering
\bigskip
\subfigure{\includegraphics[width=3.0in]{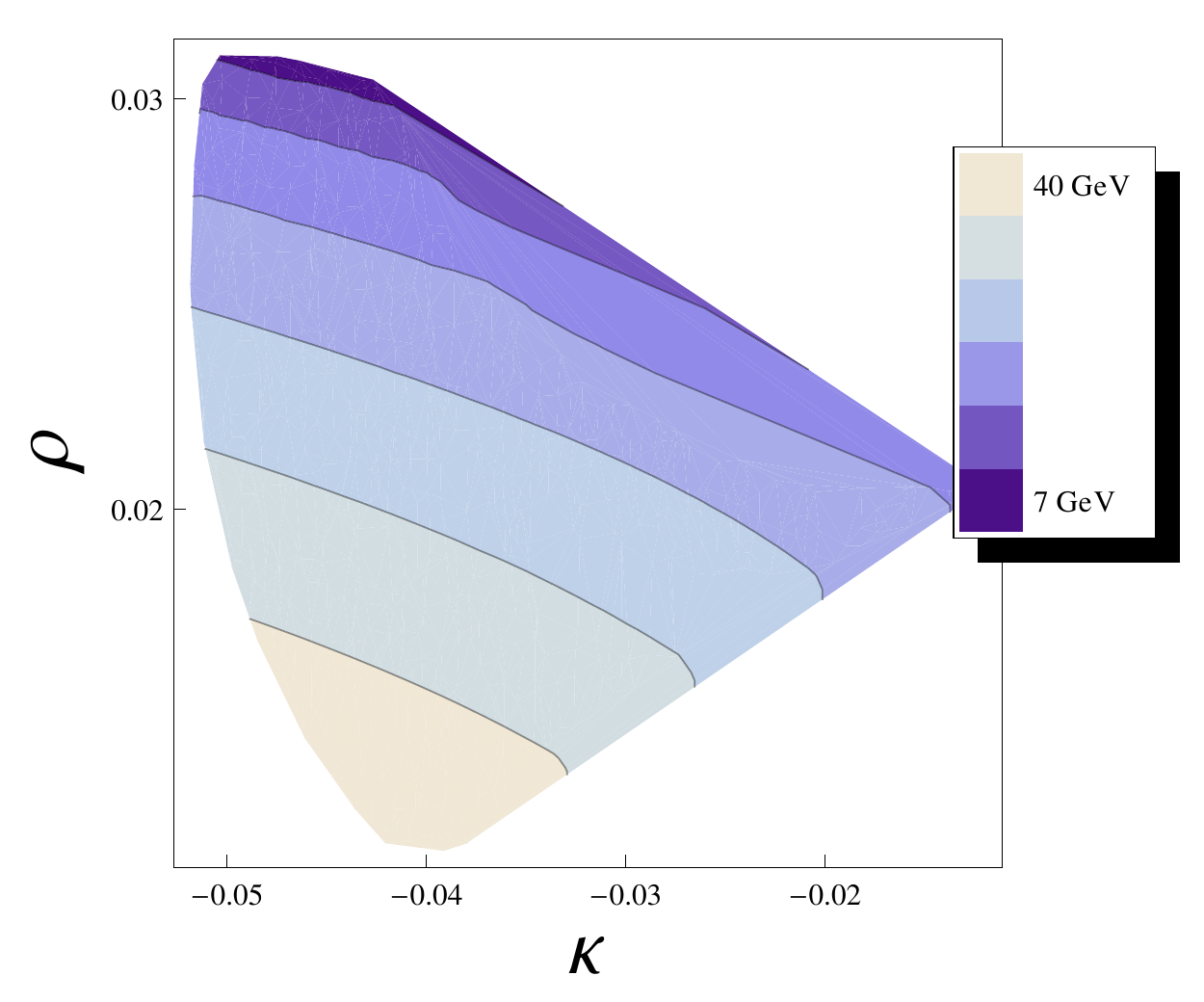}}~
\subfigure{\includegraphics[width=3.0in]{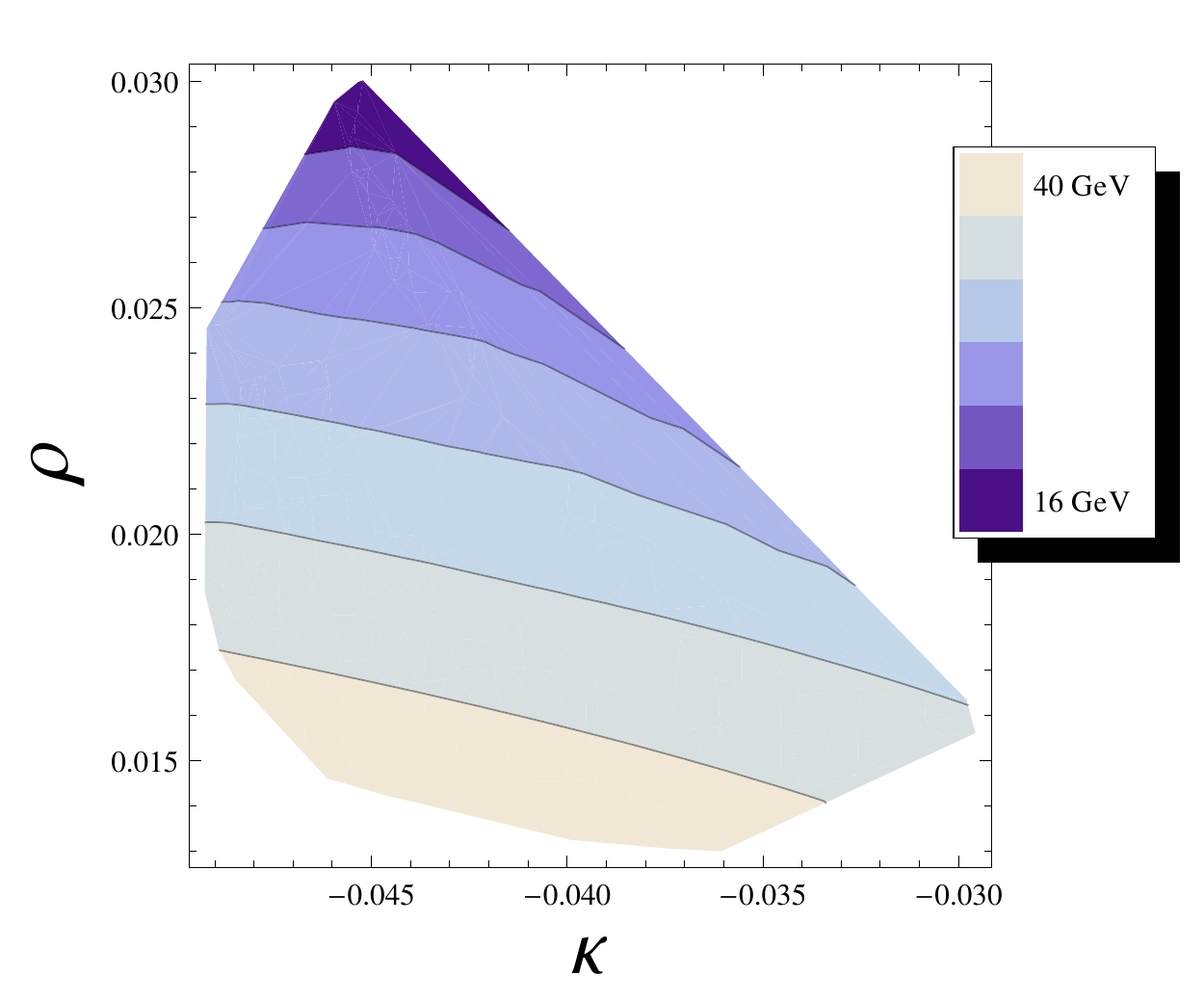}}
\caption{\small Contours of the lightest pseudoscalar mass as a function of $\kappa$ and $\rho$. On the left we show the masses after imposing that the spectrum consist of a SM-like Higgs mass of $\sim 126$ GeV and charginos consistent with LEP. On the right we show the allowed masses after all constraints are taken into account.} \label{fig:pseudo1500}
\bigskip
\end{figure}

\begin{figure}[ht]\centering
\bigskip
\subfigure{\includegraphics[width=3.0in]{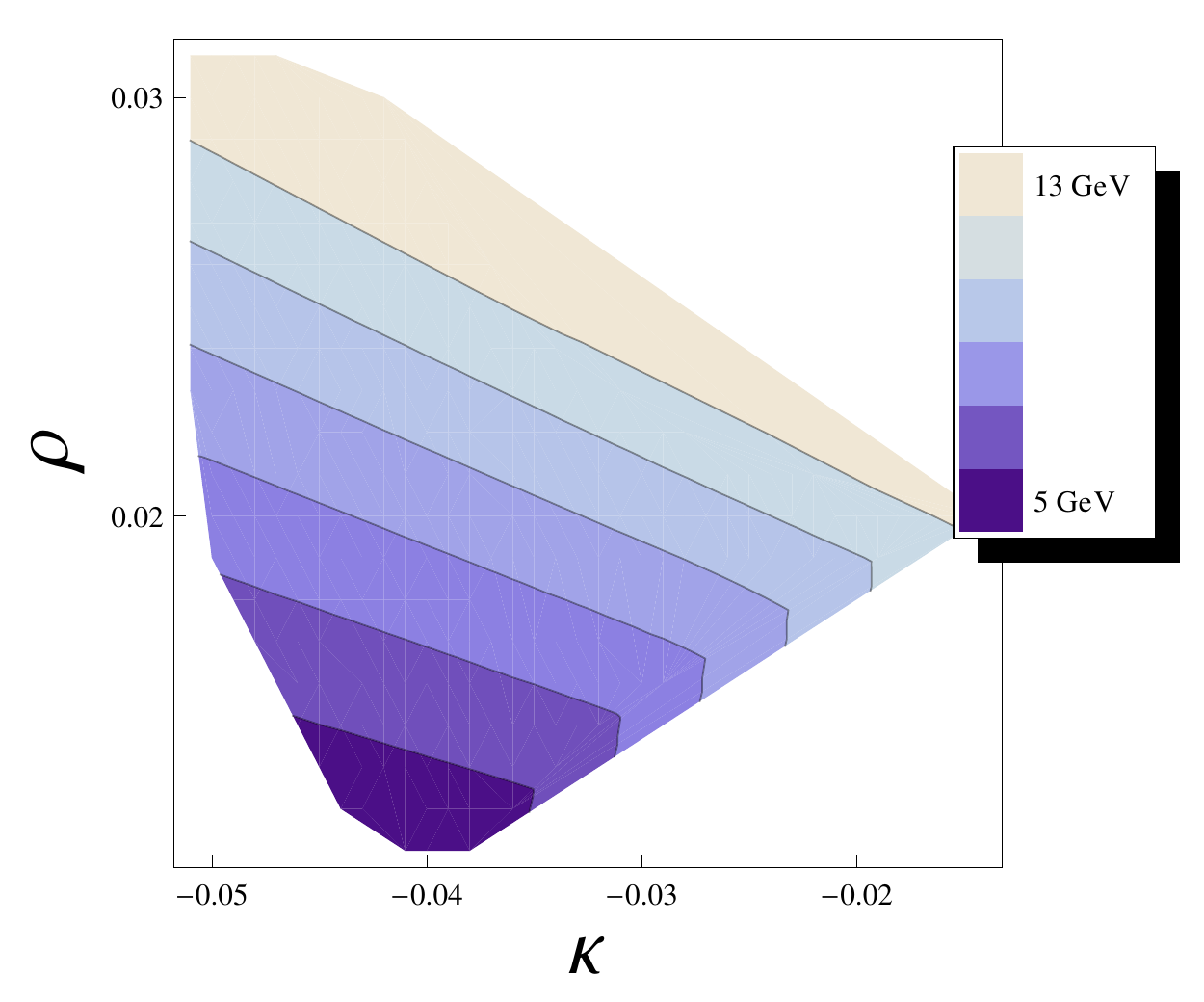}}~
\subfigure{\includegraphics[width=3.0in]{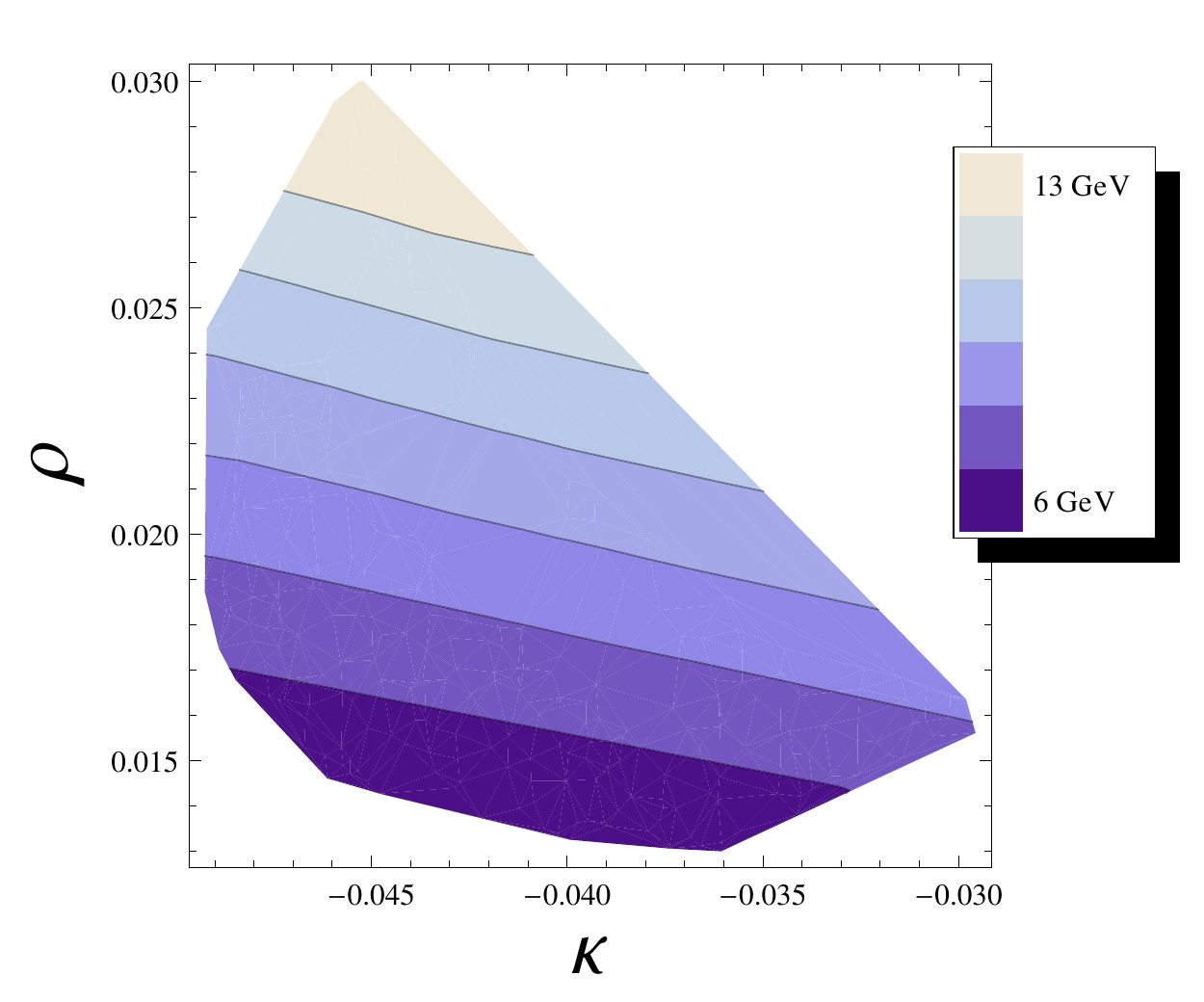}}
\caption{\small Contours of the lightest neutralino mass as a function of $\kappa$ and $\rho$. On the left we show the masses after imposing that the spectrum consist of a SM-like Higgs mass of $\sim 126$ GeV and charginos consistent with LEP. On the right we show the allowed masses after all constraints are taken into account.} \label{fig:DM1500}
\bigskip
\end{figure}

In the following section we study the cosmological abundance of a light neutralino with mass below $15$ GeV that annihilates into SM particles to produce the observed density of dark matter.
%%%%%%%%%%%%%%%%%%%%%%%%%%%%%%%%%%%%%%%%%%%%%%%%%%%%%%%%%%%%%%%%%%%%%%%%%%%%%%%%%%%%%%%%%%%%%%%%
%	
%
%
%
%%%%%%%%%%%%%%%%%%%%%%%%%%%%%%%%%%%%%%%%%%%%%%%%%%%%%%%%%%%%%%%%%%%%%%%%%%%%%%%%%%%%%%%%%%%%%%%%
\section{A Dark Matter Candidate}\label{sec:DarkMatter}

The LSP of SUSY models with exact $R-$parity is known to be a good candidate for cold dark matter \cite{Jungman:1995df}. In general, the LSP is a weekly interacting massive particle (WIMP) and, depending on the specifics of the model, it could be a neutralino, the gravitino, an sneutrino or an axino, among others. Particularly, in the context of the NMSSM, the LSP is commonly the lightest neutralino, which has a large fraction of singlino \cite{Greene:1986th,Flores:1990bt,Olive:1990aj,Abel:1992ts,Stephan:1997rv}. This favors a light DM candidate, with mass below $20$ GeV, in the PQ limit or when a continuous R-symmetry is imposed. In such class of models, the relic density is obtained through annihilation into a light scalar or a pseudoscalar Higgs boson~\cite{Gunion:2005rw,Das:2010ww,Draper:2010ew,Carena:2011jy,Kozaczuk:2013spa}. 

The abundance of thermal relics, $X$, in the universe is determined by their self-annihilation in relation to the expansion rate of the universe. In the early universe, these particles are abundant and are in thermal equilibrium with the rest of degrees of freedom. When the expansion of the universe dominates over the annihilation rate, and the universe cools down to a temperature below $m_X$, the interaction among DM particles is less efficient and their density ``freezes out". The evolution of the comoving particle density is given by the Boltzmann equation \cite{Kolb:1990vq}
\begin{equation}
\frac{dn_X}{dt}+3Hn_X\,=\,-\langle \sigma_{X\bar{X}} v \rangle \left(n_X^2-n_{X\,eq}^2\right), \label{BoltEq} 
\end{equation}
where $H$ is the Hubble rate and $\langle \sigma_{X\bar{X}} v \rangle$ is the thermally averaged annihilation cross section. 

The freeze-out temperature, $T_{FO}$, at which the particles depart from equilibrium, can be found by solving numerically equation (\ref{BoltEq}). This is, approximately,
\begin{equation}
x_{FO}\equiv \frac{m_X}{T_{FO}}\approx {\rm ln} \left(0.038 g_X \frac{m_X\,M_{Pl}\langle\sigma_{X\bar{X}} v \rangle }{g_{*}^{1/2} x_{FO}^{1/2}}\right),
\end{equation} where $g_{*}$ is the number of relativistic degrees of freedom at the freeze-out temperature. Subsequently, the present day relic abundance is given by 
\begin{equation}
\Omega_{\rm DM} h^2 \approx \frac{1.07 \times 10^9 {\rm GeV}^{-1}}{J\,g_{*}^{1/2} M_{\rm Pl}}, \qquad\, {\rm with }\qquad J\equiv \int_{x_{FO}}^\infty \frac{\langle \sigma_{X\bar{X}} v \rangle}{x^2} dx.
\end{equation} Here, $h$ is the Hubble parameter in units of $100$ ${\rm km\, s}^{-1}{\rm Mpc}^{-1}$. It is convenient to express this relic abundance in terms of the Taylor expansion of the cross section $\langle \sigma_{X\bar{X}} v \rangle \approx a + bv^2$,
\begin{equation} 
\Omega_{\rm DM} h^2 \approx \frac{1.07 \times 10^9 x_{\rm FO}}{g_{*}^{1/2} M_{\rm Pl} {\rm GeV} (a+3b/x_{\rm FO})}.
\end{equation}

We now apply this analysis to our model by considering the lightest neutralino, $\chi^0_1$, as the DM particle. In order to find its abundance, we calculate the annihilation cross section in the same fashion as it was done in \cite{Nihei:2001qs,Nihei:2002ij}, where the neutralino relic density was computed for the MSSM. As we have already mentioned, since the next-to-lightest neutralino is much heavier than the LSP, we do not include co-annihilations in our calculations.

For convenience, the function $w(s)$ is defined
\begin{equation}
w(s) \equiv \frac{1}{4}\int d {\rm LIPS} |\mathcal{M} (\chi \chi \rightarrow {\rm all})|^2 \,=\,\frac{1}{2}\sqrt{s(s-4m_\chi^2)}\sigma(s), \label{xsection}
\end{equation} where $s$ is the Mandelstam variable.

For annihilations into a two-body final state $\chi \chi \rightarrow f_1f_2$, $w(s)$ is given by 
\begin{equation}
w(s) = \frac{1}{32\pi}\sum_{\rm all} \left\lbrace c\, \theta \left( s-(m_{f_1}^2+m_{f_2}^2)^2\right) \beta_f(s, m_{f_1},m_{f_2}) \tilde{w}_{f_1f_2}(s) \right\rbrace, \label{w_s}
\end{equation} where $\theta(x)$ is the Heaviside function, $c$ is a color factor (3 a quark-antiquak final state, 1 otherwise), and 
\begin{equation}
\tilde{w}_{f_1f_2}(s) = \frac{1}{8\pi}\int d\Omega |\mathcal{M} (\chi \chi \rightarrow  f_1f_2)|^2,
\end{equation} with
\begin{equation}
\beta_f(s, m_{f_1},m_{f_2}) = \left[1-\frac{(m_{f_1}^2+m_{f_2}^2)^2}{s}\right]^{1/2}\left[1-\frac{(m_{f_1}^2-m_{f_2}^2)^2}{s}\right]^{1/2}.
\end{equation} 

Once $\sigma(s)$ is obtained, the thermally averaged cross section can be computed using
\begin{equation}
\langle \sigma_{\chi\chi} v \rangle =  \frac{1}{8 m_{\chi}^4 T K^2_2\left(m_\chi/T\right)}\int_{4m^2_\chi}^\infty \,ds \sigma(s) (s-4m_\chi^2) s^{1/2} K_1 \left(\frac{s^{1/2}}{T}\right),
\end{equation} where $K_{1,2}$ are modified Bessel functions.

The LSP in the model presented in this work is mostly singlino and very light. This implies that the kinematically allowed annihilation processes are those where the final states are light MSSM fermions. Thus, in the final state, we consider $u,\,d,\,c,\,s,\,b$ quark-antiquark pairs and lepton $\ell \bar{\ell}$-pairs. The important processes involved in the calculation of the $\chi^0_1$ relic abundance are $s$-channel annihilations through a Higgs-like scalar ($h_i$ and $A_i$) or a $Z$ boson. In the case of a CP-even scalar, $h_i$, exchange, the contribution is given by
\begin{equation}
\tilde{w}^{(h)}_{\bar{f}f} =\abs{\sum_{j=h,H,h_S,h_N}\frac{C_{S}^{ff\,j}C_{S}^{\chi \chi\,j}}{s-m_j^2+i \Gamma_j m_j} }^2(s-4m_\chi^2)(s-4m_f^2),  
\end{equation} where the couplings $C_{S}^{ff\,j}$ are obtained by inserting the mixing matrix in Equation~(\ref{neut_mass}) in the Lagrangian. The values of $C_{S}^{\chi \chi\,j}$ are given in Equation (\ref{LSPCoupl}). 

On the other hand, the (CP-odd) pseudo-scalar $A_i$ exchange yields the $s-$wave contribution 
\begin{equation}
\tilde{w}^{(A)}_{\bar{f}f} =\abs{\sum_{j=A, A_S,A_N} \frac{C_{P}^{ff\,j}C_{P}^{\chi \chi\,j}}{s-m_{j}^2+i \Gamma_{j} m_j} }^2s^2.  \label{A_ch}
\end{equation} And, finally, the Z exchange contribution is given by
\begin{align}
\tilde{w}^{(Z)}_{\bar{f}f} =&\frac{4}{3}\abs{\sum_{A_j=A, A_S,A_N} \frac{C_{A_j}^{\chi \chi\,Z}}{s-m_Z^2+i \Gamma_Z m_Z} }^2 \times \left[ 12 \abs{\sum_{j=A, A_S,A_N} \! \! C_{j}^{ff\,Z}}^2 \frac{m_\chi^2 m_f^2 (s-4m_Z^2)^2}{m_Z^2} \right. \\
&\left. +\left( \abs{C_{V}^{ff\,Z}}^2 (s+2m_f^2)+\abs{\sum_{j=A, A_S,A_N} \! \! C_{j}^{ff\,Z}}^2 (s-4m_f^2)  \right)  (s-4m_\chi^2) \right]. \nonumber
\end{align}

These results altogether give us the cross section that determines the density of $\chi^{0}_{1}$ as expressed in Equations  (\ref{xsection}) and (\ref{w_s}), where 
\begin{equation}
\tilde{w}_{\bar{f}f}=\tilde{w}^{(h,H)}_{\bar{f}f}+\tilde{w}^{(A)}_{\bar{f}f} +\tilde{w}^{(Z)}_{\bar{f}f}.
\end{equation}

We explore what values of the parameters in our model yield a relic abundance that is consistent with the measured DM density, which corresponds to a thermally averaged annihilation cross section $\langle \sigma v \rangle \sim 3\times 10^{-26} {\rm cm}^3/{\rm s}$. To do so, we scan the parameter space over the ranges presented in Table~\ref{tab:TabParam2} and impose the invisible $Z$ decay constraints. Also, we require a realistic Higgs mass, $m_h \approx 126$ GeV, and that $m_{\chi^+} >104$ GeV to be consistent with collider results.

Our findings show that a lightest neutralino with mass between $4$ GeV and $9$ GeV yields the appropriate relic density, as shown in Figure \ref{fig:CrossSect_a}. for this mass range, there is a significant component of $\chi^0_1$ along $\tilde{H}_u$, as depicted in Figure \ref{fig:CrossSect_b}. Additionally, the annihilation is dominated by the (s-wave) interchange of a light CP-odd scalar, given in Equation (\ref{A_ch}), and as depicted in Figure \ref{fig:CrossSecPseudoRatio}, while the contribution from the CP-even scalar mediated annihilation is p-wave suppressed. Therefore, the presence of light pseudoscalars in this model aides in making the annihilation of the relics efficient, avoiding an overabundance of the DM particles. The DM mass obtained in this analysis is significantly smaller than most MSSM neutralino-like proposals, and it is also consistent with the studies of light DM in the NMSSM  \cite{Gunion:2005rw,Das:2010ww,Draper:2010ew,Carena:2011jy,Kozaczuk:2013spa}.

\begin{figure}[ht]\centering
\bigskip
\subfigure{\includegraphics[height=1.9in,width=2.8in]{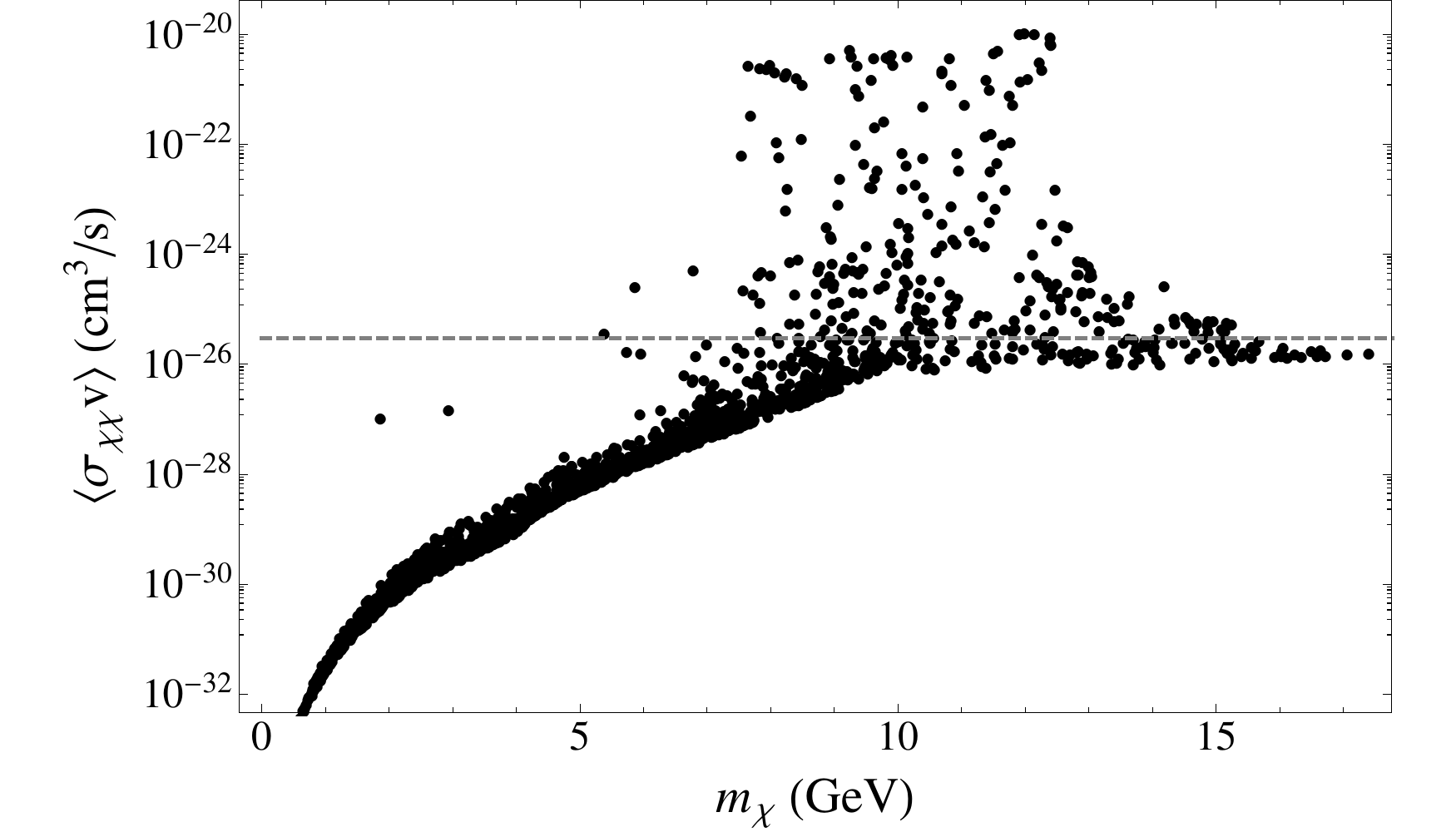}\label{fig:CrossSect_a}}~
\subfigure{\includegraphics[width=2.8in]{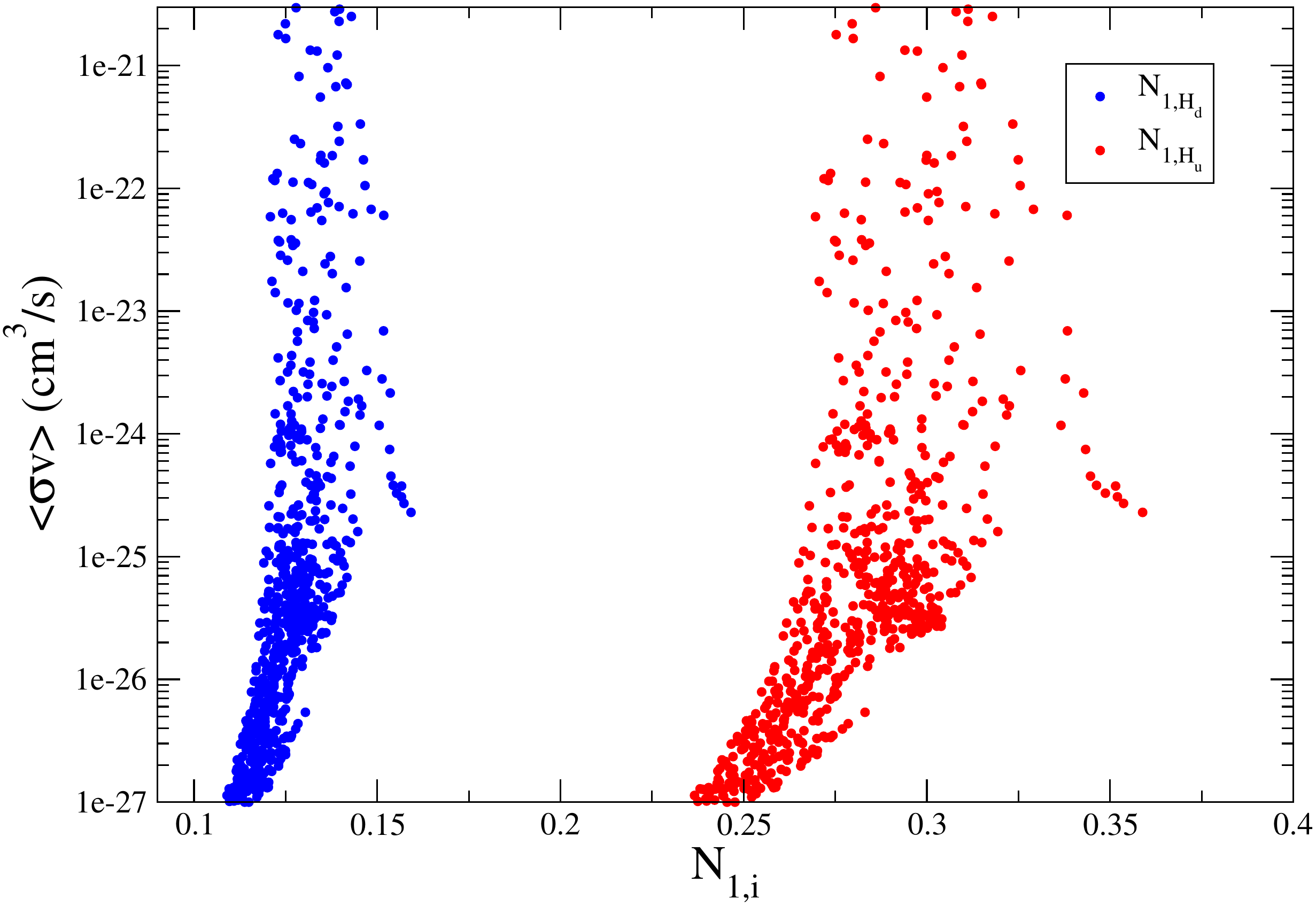} \label{fig:CrossSect_b}}
\caption{\small Annihilation cross section for $\chi^0_1$. Figure (a) on the left shows the cross section as a function of the neutralino mass, whereas panel (b) on the right depicts the mixing components of the $\chi^0_1$.}
\end{figure}

\begin{figure}[ht]\centering
\subfigure{\includegraphics[width=2.8in]{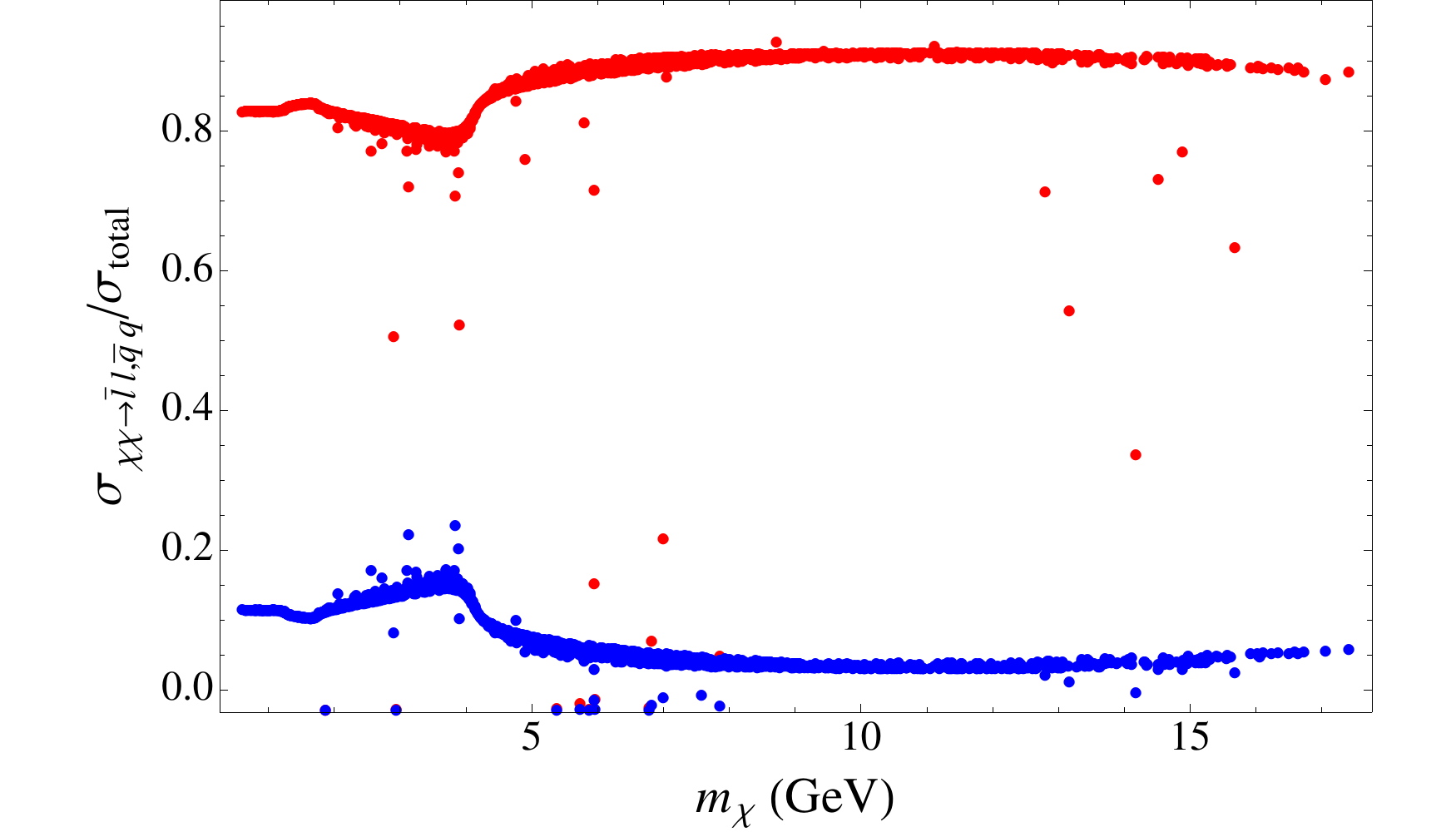}\label{fig:CrossSecProdRatio}}~
\subfigure{\includegraphics[width=2.8in]{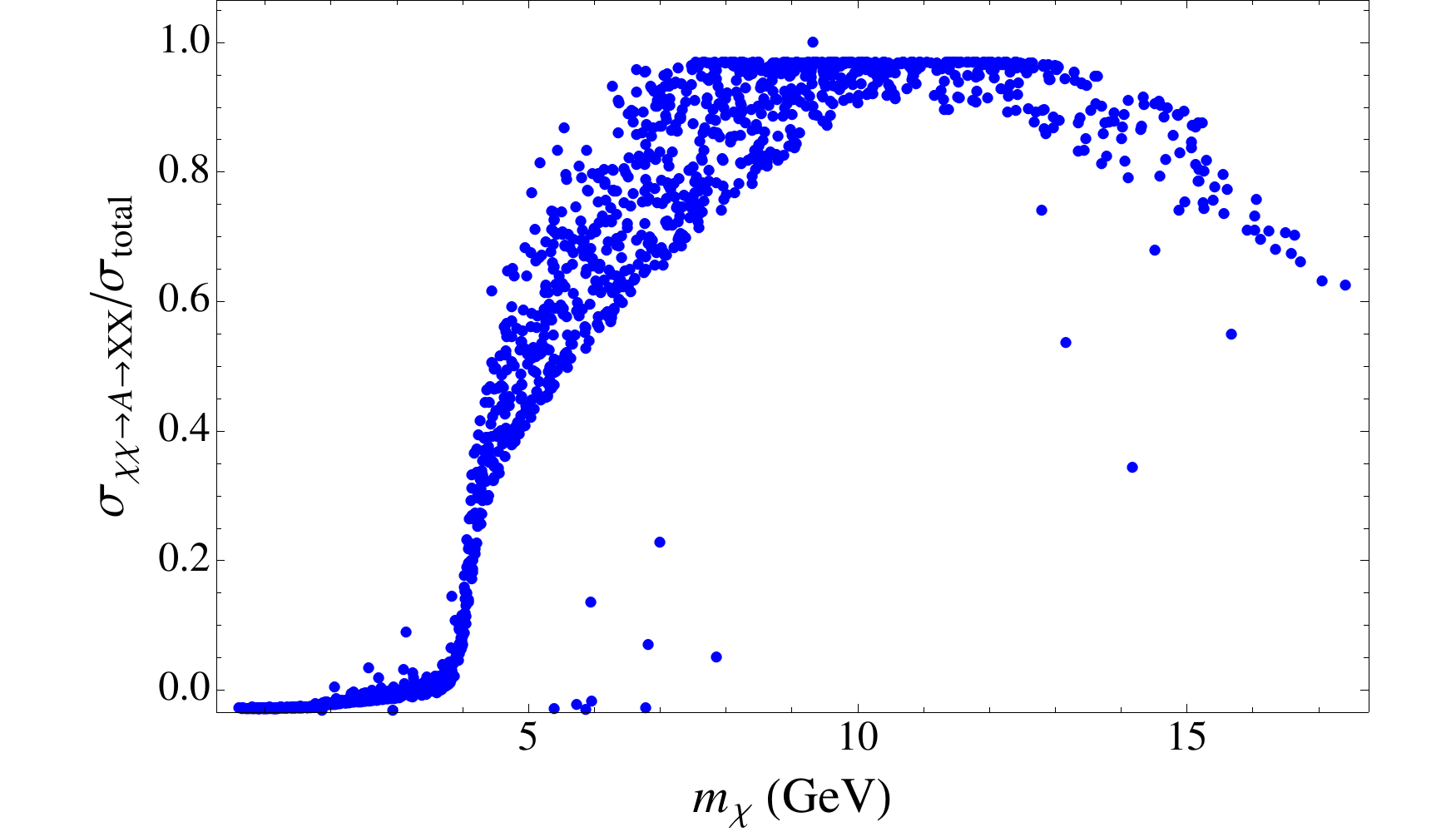} \label{fig:CrossSecPseudoRatio}}
\caption{\small Figure (a) on the left shows the ratio of neutralino annihilation cross-section to leptons (blue) and quarks (red). Figure (b) on the right depicts ratio of CP-odd scalar channel neutralino annihilation cross-section.}
\end{figure}

The final products from the annihilation of this light neutralino are $\bar{\ell}\ell$ pairs or light $\bar{q}q$ pairs. This is shown in Figure \ref{fig:CrossSecProdRatio}, where the dominant process is that of annihilation into a pair of quarks, specially for neutralino masses close to 10 GeV. In particular, the dominant process for $m_{\chi^0_1} \gtrsim 4\,{\rm GeV}$ yields a $\bar{b} b$ pair final state, whereas below this mass the most relevant products are $\bar{d}d,\,\bar{u}u$ and $\bar{s}s$ pairs.

Let us now take a look at the effects of the wino mass, $M_{2}$, on the allowed neutralino mass values consistent with a thermalized cross section of $3\times 10^{-26}$ cm$^{3}$/s. For $M_{2}=500$ GeV, it was shown in the previous section that after all constraints are taken into consideration, the values of $\kappa$ and $\rho$ are highly restricted. In particular, only values of $\rho\sim 0.023$ are allowed. This yields a light neutralino and light pseudoscalar mass of $8$ and $30$ GeV respectively. In this benchmark scenario the annihilation cross section is not resonant for $2m_{\chi^{0}_{1}}\approx m_{A_{N}}$, and thermalized cross sections above $1.0 \times 10^{-26}\text{cm}^{3}/\text{s}$ are not viable. The situation is different for $M_{2}=1.5$ TeV, where the values of $\kappa$ and $\rho$ are less restricted. This can be seen in Figures~\ref{fig:CrossNeut} and~\ref{fig:CrossPseudo}, where we show the annihilation cross section as a function of the lightest neutralino and pseudoscalar masses respectively. From the figure, one can see that the annihilation is most efficient when the pseudoscalar mediator is light or when the lightest neutralino and pseudoscalar satisfy the resonant condition $2m_{\chi^{0}_{1}}\approx m_{A_{N}}$.

\begin{figure}[ht]\centering
\bigskip
\subfigure{\includegraphics[width=2.8in]{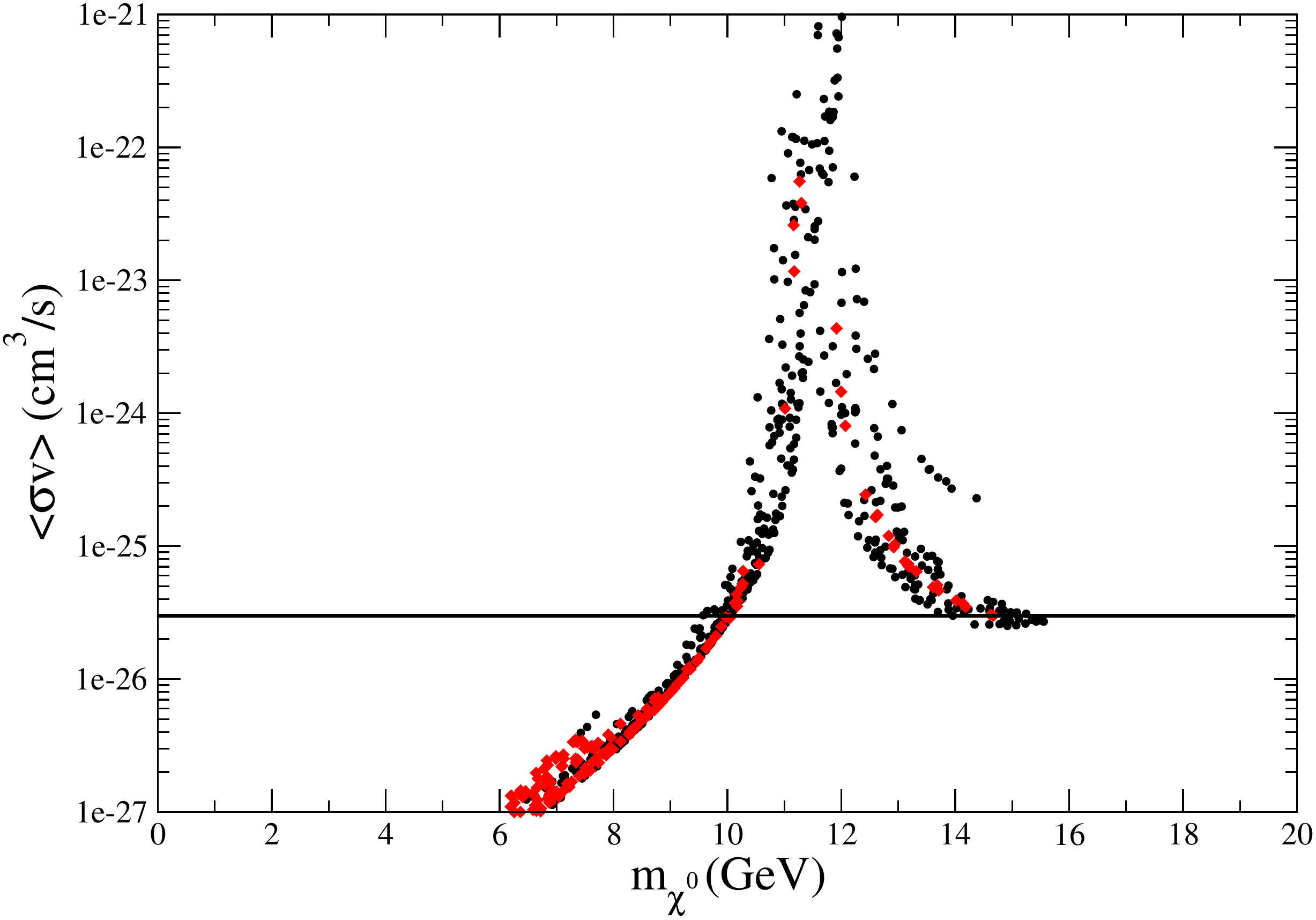}\label{fig:CrossNeut}}~
\subfigure{\includegraphics[width=2.8in]{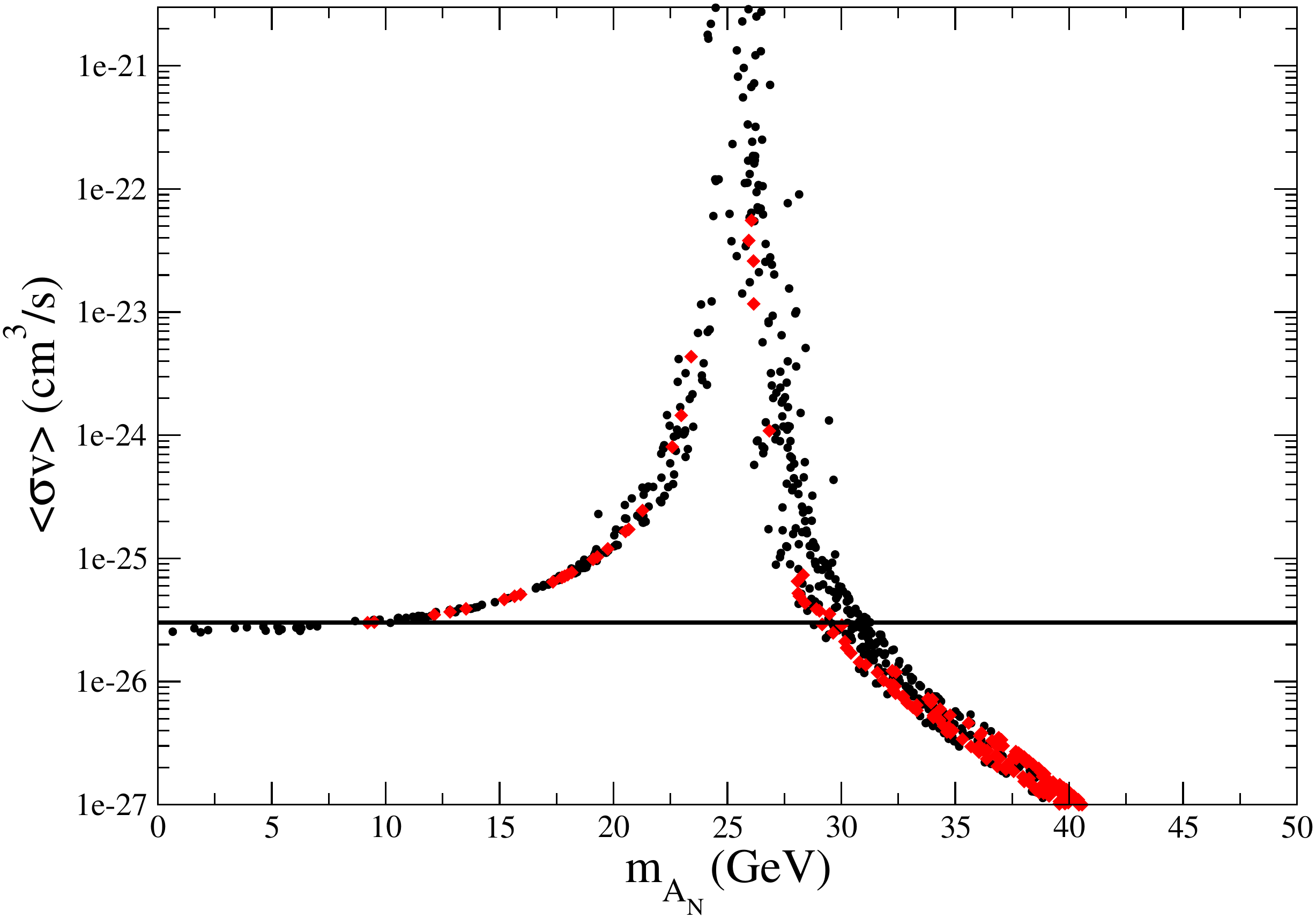} \label{fig:CrossPseudo}}
\caption{\small Annihilation cross section as function of the lightest neutralino mass on the left, Figure (a), and the lightest pseudoscalar mass on the right, Figure (b). The black dots are points which are consistent with a $126$ GeV SM-like Higgs, the invisible decay width of the $Z$ and the the chargino mass bound while the red dots are consistent with all of the constraints introduced in Section~\ref{sec:Constraints}. The balck solid line corresponds to a value of $\left<\sigma v\right>=3\times 10^{-26}$cm$^{3}$/s. }
\end{figure}

Finally, a comment about the detection possibilities for this scenario is in order. The dominant singlino nature of the DM particle in our model makes it significantly decoupled from the MSSM degrees of freedom. The spin independent elastic scattering cross section of $\chi^0_1$ with nucleons is given by 
\be 
\sigma_{SI}^{p,n} = \sum_{H}\frac{1}{m_H^4} \left(\frac{m_{p,n} m_{\tilde{\chi}_0}}{m_p+m_{\tilde{\chi}_0}}\right)^2( C_{S}^{\chi \chi H})^2 \left( \sum_{q} C_{q\bar{q}H} \langle N | q\bar{q} |N\rangle \right)^2,
\ee which in our case yields a value of the order of $10^{-48}-10^{-46} $ ${\rm cm}^2$ for $m_{\chi^0_1}\approx 10$ GeV. This is below the range of cross sections that the current direct detection experiments are able to measure, $\sigma_{SI} \sim 10^{-46} \,{\rm cm}^2$ \cite{Bauer:2013ihz}. However, this cross section sensitivity might be achieved in future detectors. Notice that cross sections of this magnitude are just below the values for which the irreducible neutrino background would affect the discrimination capabilities of the detector \cite{Gutlein:2010tq}.  Despite the fact that in our model the DM particle has a mass that is close to the mass hinted by signals detected recently in CDMS \cite{Agnese:2013rvf}, where the DM-nucleon scattering cross section in their detected events is about $10^{-41}$ ${\rm cm}^2$, which is far from the expected cross section in our model. Recently, the LUX experiment has released the results of their first WIMP search \cite{Akerib:2013tjd}. They find an upper bound for the annihilation cross section of $7.6\times 10^{-46}\, {\rm cm}^2$ for a WIMP mass of $30$ GeV. The limits corresponding to the range of masses considered in this work are between $10^{-44}\, {\rm cm}^2$ and $10^{-45}\, {\rm cm}^2$.

 %%%%%%%%%%%%%%%%%%%%%%%%%%%%%%%%%%%%%%%%%%%%%%%%%%%%%%%%%%%%%%%%%%

 %%%%%%%%%%%%%%%%%%%%%%%%%%%%%%%%%%%%%%%%%%%%%%%%%%%%%%%%%%%%%%%%%%
%
%
%%%%%%%%%%%%%%%%%%%%%%%%%%%%%%%%%%%%%%%%%%%%%%%%%%%%%%%%%%%%%%%%%%%
\section{Conclusions}\label{sec:Conclusions}

Extensions of the MSSM have been extensively used in the literature to solve the $\mu$ and little hierarchy problems. In this article, we have explored the Higgs and neutralino sectors for an extension of the MSSM, in which those problems are easily addressed. We performed a scan of the parameter space and found the regions that are consistent with collider constraints and a Higgs mass around $126$ GeV. In the Higgs sector, we have found two singlet-like scalars that are allowed by present constraints. In the neutralino sector, we have investigated the existence of a light dark matter candidate and its annihilation cross section. In fact, the dark matter particle is ``mostly" the fermionic partner of a singlet scalar that does not couple directly to the ordinary matter, but only through a small coupling to the usual singlet present in the NMSSM. 

This relic particle turns out to have a mass in the range $8\, {\rm GeV}\,<\, m_{\chi}\,< \,15 \,{\rm GeV}$, which is considerably lighter than candidates for dark matter in the MSSM. Its interaction is also remarkably weak, more than the expected interaction in the usual WIMP scenarios. However, the presence of the new singlet-like scalars, and specially the lightest pseudoscalar, favors the annihilation process, and the right relic abundance can be obtained for a wide region of the parameter space. This provides an example of a scenario where the dark matter is somewhat hidden, with the singlet field $S$ field acting as a portal to the MSSM matter content. Along these lines, we found that the cross section sensitivities of the current direct detection experiments are just above the estimated scattering cross section of this dark matter particle with the nucleons, which makes the detection of this type of relic unachievable at present, but it could be tested in future experiments.

Finally, this model has been studied at the phenomenological level; it would be interesting to explore the completion at high energies such as embedding this construction in a gauge mediated SUSY breaking scenario, similar to that presented in \cite{Delgado:2010cw} for the S-MSSM.

 %%%%%%%%%%%%%%%%%%%%%%%%%%%%%%%%%%%%%%%%%%%%%%%%%%%%%%%%%%%%%%%%%%
%
%
%%%%%%%%%%%%%%%%%%%%%%%%%%%%%%%%%%%%%%%%%%%%%%%%%%%%%%%%%%%%%%%%%%%
\section*{Acknowledgements}

ADP would like to thank David Morrissey, Antonio Delgado and Jorge de Blas Mateo for useful discussions and essential feedback regarding the progress of this work. WT is grateful to Can Kilic and and Jiang-Hao Yu for very useful discussions. The research of ADP is supported by the National Science and Engineering Research Council. WT is supported by the National Science Foundation under Grant PHY-0969020, the Texas Cosmology Center and by a University of Texas continuing fellowship.

%%%%%%%%%%%%%%%%%%%%%%%%%%%%%%%%%%%%%%%%%%%%%%%%%%%%%%%%%%%%%%%%%%%
%	
%									Appendix
%
%
%%%%%%%%%%%%%%%%%%%%%%%%%%%%%%%%%%%%%%%%%%%%%%%%%%%%%%%%%%%%%%%%%%%

\appendix

%%%%%%%%%%%%
\section{Renormalization Group Equations}
\label{sec:rges}
%%%%%%%%%%%%
In this appendix we give the renormalization group equations to one loop order using the conventions found in~\cite{Martin:1997ns}:
\begin{equation}
\beta_{y^{ijk}}=\frac{dy^{ijk}}{dt}=\gamma^{i}_{n}y^{njk}+\gamma^{j}_{n}y^{ink}+\gamma^{k}_{n}y^{ijn}.
\end{equation}
In what follows, we include the gauge couplings, all dimensionless superpotential scalar couplings and the Yukawa couplings for the third family. The conventions used are $t=\log{\mu/\text{GeV}}$ and a $U(1)_{Y}$ gauge coupling related to the $SU(5)$ normalization by $g_{1}=\sqrt{\frac{3}{5}}g^{SU(5)}_{1}$.

\begin{eqnarray}
   g_3'(t)&=&-\frac{3 g_3{}^3}{16 \pi ^2}, \nonumber \\
   g_2'(t)&=&\frac{g_2{}^3}{16 \pi ^2}, \nonumber \\
   g_1'(t)&=&\frac{33 g_1{}^3}{80 \pi ^2}, \nonumber \\
   y_1'(t)&=&\frac{y_1 \left(-\frac{13}{15} g_1{}^2-3 g_2{}^2-\frac{16}{3}
   g_3{}^2+\lambda^2+6 y_1{}^2+y_2{}^2\right)}{16 \pi ^2}, \nonumber \\
   y_2'(t)&=&\frac{y_2 \left(-\frac{7}{15} g_1{}^2-3 g_2{}^2-\frac{16}{3}
   g_3{}^2+\lambda^2+y_1{}^2+6 y_2{}^2+y_3{}^2\right)}{16 \pi ^2}, \nonumber \\
   y_3'(t)&=&\frac{y_3 \left(-\frac{9}{5} g_1{}^2-3 g_2{}^2+\lambda^2+3
   y_2{}^2+4 y_3{}^2\right)}{16 \pi ^2}, \nonumber \\
   \lambda '(t)&=&\frac{\lambda \left(-\frac{3}{5} g_1{}^2-3 g_2{}^2+2 \alpha
   _1{}^2+2 \kappa^2+4 \lambda^2+2\lambda^{2}_{N}+4 \rho^2+3 y_1{}^2+3
   y_2{}^2+y_3{}^2\right)}{16 \pi ^2} \nonumber \\
   &+&\frac{\lambda_{N}\left(2\lambda\lambda_{N}+2\kappa\rho+4\rho\alpha_{1}+2\alpha_{1}\alpha_{2}\right)}{16\pi^{2}}, \nonumber \\
   \lambda_{N} '(t)&=&\frac{\lambda_{N} \left(-\frac{3}{5} g_1{}^2-3 g_2{}^2+4 \alpha
   _1{}^2+2\alpha^{2}_{2}+4 \lambda_{N}^2+2\lambda^{2}+2 \rho^2+3 y_1{}^2+3
   y_2{}^2+y_3{}^2\right)}{16 \pi ^2} \nonumber \\
   &+&\frac{\lambda\left(2\lambda\lambda_{N}+2\kappa\rho+4\rho\alpha_{1}+2\alpha_{1}\alpha_{2}\right)}{16\pi^{2}}, \nonumber \\
   \kappa '(t)&=&\frac{\kappa  \left(6 \alpha _1{}^2+6 \kappa ^2+6 \lambda ^2+12
   \rho ^2\right)}{16 \pi ^2} + \frac{\rho  \left(6\lambda\lambda_{N}+12 \alpha _1 \rho +6 \alpha
   _2{}^2+6 \kappa  \rho \right)}{16 \pi ^2}, \nonumber \\
   \rho '(t)&=&\frac{\rho  \left(8 \alpha _1{}^2+2 \alpha _2{}^2+4 \kappa ^2+4
   \lambda ^2+10 \rho ^2+2\lambda^{2}_{N}\right)}{16 \pi ^2} + \frac{\alpha _1 \left(4\lambda\lambda_{N}+8 \alpha
   _1 \rho +4 \alpha _1 \alpha _2+4 \kappa \rho \right)}{16 \pi ^2} \nonumber \\
   &+&\frac{\kappa\left(2\lambda\lambda_{N}+2\kappa\rho+4\rho\alpha_{1}+2\alpha_{1}\alpha_{2}\right)}{16\pi^{2}}, \nonumber \\
   \alpha _1'(t)&=&\frac{\alpha _1 \left(10 \alpha _1{}^2+4 \alpha _2{}^2+2 \kappa
   ^2+2 \lambda ^2+4\lambda^{2}_{N}+8 \rho ^2\right)}{16 \pi ^2} +\frac{\alpha _2 \left(2\lambda\lambda_{N}+4
   \alpha _1 \rho +2 \alpha _1 \alpha _2+2 \kappa \rho\right)}{16
   \pi ^2} \nonumber \\
   &+&\frac{\rho  \left(4\lambda\lambda_{N}+8 \alpha _1 \rho+4 \alpha _1 \alpha _2+4
   \kappa  \rho \right)}{16 \pi ^2}, \nonumber \\
   \alpha _2'(t)&=&\frac{\alpha _1 \left(6\lambda\lambda_{N}+12 \alpha _2 \rho +6 \alpha _1 \alpha
   _2+6 \kappa  \rho \right)}{16 \pi ^2} +\frac{\alpha _2 \left(6\lambda^{2}_{N}+12 \alpha
   _1{}^2+6 \alpha _2{}^2+6 \rho ^2\right)}{16 \pi ^2},
   \end{eqnarray}
   where $y_{1}$ denotes the Yukawa coupling for the $top$-quark and $y_{2},y_{b}$ for the $bottom$ quark and $tau$ lepton respectively.

%%%%%%%%%%%%%%%%%%%%%%%%%%%%%%%%%%%%%%%%%%%%%%%%%%%%%%%%%%%%%%%%%%%
%	
%									REFERENCES
%
%
%%%%%%%%%%%%%%%%%%%%%%%%%%%%%%%%%%%%%%%%%%%%%%%%%%%%%%%%%%%%%%%%%%%%%%%%%%%%%


\begin{thebibliography}{99}


%\cite{Barate:2003sz}
\bibitem{Barate:2003sz} 
  R.~Barate {\it et al.}  [LEP Working Group for Higgs boson searches and ALEPH and DELPHI and L3 and OPAL Collaborations],
  ``Search for the standard model Higgs boson at LEP,''
  Phys.\ Lett.\ B {\bf 565}, 61 (2003)
  [hep-ex/0306033].
  %%CITATION = HEP-EX/0306033;%%
  %1897 citations counted in INSPIRE as of 12 Apr 2013
  
  %\cite{Aad:2012tfa}
\bibitem{Aad:2012tfa} 
  G.~Aad {\it et al.}  [ATLAS Collaboration],
  ``Observation of a new particle in the search for the Standard Model Higgs boson with the ATLAS detector at the LHC,''
  Phys.\ Lett.\ B {\bf 716}, 1 (2012)
  [arXiv:1207.7214 [hep-ex]].
  %%CITATION = ARXIV:1207.7214;%%
  %908 citations counted in INSPIRE as of 12 Apr 2013

%\cite{Chatrchyan:2013lba}
\bibitem{Chatrchyan:2013lba} 
  S.~Chatrchyan {\it et al.}  [ CMS Collaboration],
  ``Observation of a new boson with mass near 125 GeV in pp collisions at sqrt(s) = 7 and 8 TeV,''
  arXiv:1303.4571 [hep-ex].
  %%CITATION = ARXIV:1303.4571;%%
  %4 citations counted in INSPIRE as of 12 Apr 2013
  
  
    %\cite{Draper:2011aa}
\bibitem{Draper:2011aa} 
  P.~Draper, P.~Meade, M.~Reece and D.~Shih,
  ``Implications of a 125 GeV Higgs for the MSSM and Low-Scale SUSY Breaking,''
Phys.\ Rev.\ D {\bf 85}, 095007 (2012)
[arXiv:1112.3068 [hep-ph]].
%%CITATION = ARXIV:1112.3068;%%
  %154 citations counted in INSPIRE as of 19 Sep 2013

%\cite{Carena:2011aa}
\bibitem{Carena:2011aa} 
  M.~Carena, S.~Gori, N.~R.~Shah and C.~E.~M.~Wagner,
  ``A 125 GeV SM-like Higgs in the MSSM and the $\gamma \gamma$ rate,''
JHEP {\bf 1203}, 014 (2012)
[arXiv:1112.3336 [hep-ph]].
%%CITATION = ARXIV:1112.3336;%%
  %229 citations counted in INSPIRE as of 19 Sep 2013
  
  
%\cite{Dimopoulos:1995mi}
\bibitem{Dimopoulos:1995mi} 
  S.~Dimopoulos and G.~F.~Giudice,
  ``Naturalness constraints in supersymmetric theories with nonuniversal soft terms,''
  Phys.\ Lett.\ B {\bf 357}, 573 (1995)
  [hep-ph/9507282].
  %%CITATION = HEP-PH/9507282;%%
  
  
    %\cite{Barbieri:2000gf}
\bibitem{Barbieri:2000gf} 
  R.~Barbieri and A.~Strumia,
  ``The 'LEP paradox',''
  hep-ph/0007265.
  %%CITATION = HEP-PH/0007265;%%
  
  %\cite{Kitano:2006gv}
\bibitem{Kitano:2006gv} 
  R.~Kitano and Y.~Nomura,
  ``Supersymmetry, naturalness, and signatures at the LHC,''
  Phys.\ Rev.\ D {\bf 73}, 095004 (2006)
  [hep-ph/0602096].
  %%CITATION = HEP-PH/0602096;%%
  
  %\cite{Espinosa:1992hp}
\bibitem{Espinosa:1992hp} 
  J.~R.~Espinosa and M.~Quiros,
  ``Upper bounds on the lightest Higgs boson mass in general supersymmetric Standard Models,''
  Phys.\ Lett.\ B {\bf 302}, 51 (1993)
  [hep-ph/9212305].
  %%CITATION = HEP-PH/9212305;%%
  %170 citations counted in INSPIRE as of 12 Apr 2013

%\cite{Kane:1992kq}
\bibitem{Kane:1992kq} 
  G.~L.~Kane, C.~F.~Kolda and J.~D.~Wells,
  ``Calculable upper limit on the mass of the lightest Higgs boson in any perturbatively valid supersymmetric theory,''
  Phys.\ Rev.\ Lett.\  {\bf 70}, 2686 (1993)
  [hep-ph/9210242].
  %%CITATION = HEP-PH/9210242;%%
  %227 citations counted in INSPIRE as of 12 Apr 2013


%\cite{Maniatis:2009re}
\bibitem{Maniatis:2009re} 
  M.~Maniatis,
  %``The Next-to-Minimal Supersymmetric extension of the Standard Model reviewed,''
  Int.\ J.\ Mod.\ Phys.\ A {\bf 25}, 3505 (2010)
  [arXiv:0906.0777 [hep-ph]].
  %%CITATION = ARXIV:0906.0777;%%

%\cite{Ellwanger:2008py}
\bibitem{Ellwanger:2008py} 
  U.~Ellwanger, C.~-C.~Jean-Louis and A.~M.~Teixeira,
  ``Phenomenology of the General NMSSM with Gauge Mediated Supersymmetry Breaking,''
  JHEP {\bf 0805}, 044 (2008)
  [arXiv:0803.2962 [hep-ph]].
  %%CITATION = ARXIV:0803.2962;%%
  %28 citations counted in INSPIRE as of 12 Apr 2013

%\cite{Martin:1997ns}
\bibitem{Martin:1997ns} 
  S.~P.~Martin,
  ``A Supersymmetry primer,''
  In *Kane, G.L. (ed.): Perspectives on supersymmetry II* 1-153
  [hep-ph/9709356].
  %%CITATION = HEP-PH/9709356;%%
  %1647 citations counted in INSPIRE as of 12 Apr 2013
  
  %\cite{Ellwanger:2006rm}
\bibitem{Ellwanger:2006rm} 
  U.~Ellwanger and C.~Hugonie,
  ``The Upper bound on the lightest Higgs mass in the NMSSM revisited,''
  Mod.\ Phys.\ Lett.\ A {\bf 22}, 1581 (2007)
  [hep-ph/0612133].
  %%CITATION = HEP-PH/0612133;%%
  %44 citations counted in INSPIRE as of 15 Jul 2013
  
    %\cite{Abel:1995wk}
\bibitem{Abel:1995wk} 
  S.~A.~Abel, S.~Sarkar and P.~L.~White,
  ``On the cosmological domain wall problem for the minimally extended supersymmetric standard model,''
  Nucl.\ Phys.\ B {\bf 454}, 663 (1995)
  [hep-ph/9506359].
  %%CITATION = HEP-PH/9506359;%%
  %159 citations counted in INSPIRE as of 24 Jun 2013
  
  %\cite{Panagiotakopoulos:1998yw}
\bibitem{Panagiotakopoulos:1998yw} 
  C.~Panagiotakopoulos and K.~Tamvakis,
  ``Stabilized NMSSM without domain walls,''
  Phys.\ Lett.\ B {\bf 446}, 224 (1999)
  [hep-ph/9809475].
  %%CITATION = HEP-PH/9809475;%%
  %137 citations counted in INSPIRE as of 15 Jul 2013
  
  %\cite{Panagiotakopoulos:1999ah}
\bibitem{Panagiotakopoulos:1999ah} 
  C.~Panagiotakopoulos and K.~Tamvakis,
  ``New minimal extension of MSSM,''
  Phys.\ Lett.\ B {\bf 469}, 145 (1999)
  [hep-ph/9908351].
  %%CITATION = HEP-PH/9908351;%%
  %104 citations counted in INSPIRE as of 15 Jul 2013
  
  %%%%%%%%%%%%%%%%%%%%   125 GeV Higgs in NMSSM      %%%%%%%%%%%%%%%%%%%%%%%%%%%%%
  
  %\cite{King:2012is}
\bibitem{King:2012is}
  S.~F.~King, M.~Muhlleitner and R.~Nevzorov,
  ``NMSSM Higgs Benchmarks Near 125 GeV,''
  Nucl.\ Phys.\ B {\bf 860}, 207 (2012)
  [arXiv:1201.2671 [hep-ph]].
  %%CITATION = ARXIV:1201.2671;%%
  %122 citations counted in INSPIRE as of 12 May 2014


%\cite{Gunion:2012zd}
\bibitem{Gunion:2012zd}
  J.~F.~Gunion, Y.~Jiang and S.~Kraml,
  ``The Constrained NMSSM and Higgs near 125 GeV,''
  Phys.\ Lett.\ B {\bf 710}, 454 (2012)
  [arXiv:1201.0982 [hep-ph]].
  %%CITATION = ARXIV:1201.0982;%%
  %105 citations counted in INSPIRE as of 12 May 2014
  
  %\cite{Ellwanger:2012ke}
\bibitem{Ellwanger:2012ke}
  U.~Ellwanger and C.~Hugonie,
 ``Higgs bosons near 125 GeV in the NMSSM with constraints at the GUT scale,''
  Adv.\ High Energy Phys.\  {\bf 2012}, 625389 (2012)
  [arXiv:1203.5048 [hep-ph]].
  %%CITATION = ARXIV:1203.5048;%%
  %81 citations counted in INSPIRE as of 12 May 2014
  
  %\cite{Vasquez:2012hn}
\bibitem{Vasquez:2012hn}
  D.~A.~Vasquez, G.~Belanger, C.~Boehm, J.~Da Silva, P.~Richardson and C.~Wymant,
  ``The 125 GeV Higgs in the NMSSM in light of LHC results and astrophysics constraints,''
  Phys.\ Rev.\ D {\bf 86}, 035023 (2012)
  [arXiv:1203.3446 [hep-ph]].
  %%CITATION = ARXIV:1203.3446;%%
  %68 citations counted in INSPIRE as of 12 May 2014
  
  %\cite{Gunion:2012gc}
\bibitem{Gunion:2012gc}
  J.~F.~Gunion, Y.~Jiang and S.~Kraml,
  ``Could two NMSSM Higgs bosons be present near 125 GeV?,''
  Phys.\ Rev.\ D {\bf 86}, 071702 (2012)
  [arXiv:1207.1545 [hep-ph]].
  %%CITATION = ARXIV:1207.1545;%%
  %65 citations counted in INSPIRE as of 12 May 2014
  
  
  
    %\cite{Bae:2012am}
\bibitem{Bae:2012am} 
  K.~J.~Bae, K.~Choi, E.~J.~Chun, S.~H.~Im, C.~B.~Park and C.~S.~Shin,
  ``Peccei-Quinn NMSSM in the light of 125 GeV Higgs,''
  JHEP {\bf 1211}, 118 (2012)
  [arXiv:1208.2555 [hep-ph]].
  %%CITATION = ARXIV:1208.2555;%%
  
  
    \bibitem{Agashe125gev} 
  K.~Agashe, Y.~Cui and R.~Franceschini,
  ``Natural Islands for a 125 GeV Higgs in the scale-invariant NMSSM,''
  JHEP {\bf 1302}, 031 (2013)
  [arXiv:1209.2115 [hep-ph]].
  %%CITATION = ARXIV:1209.2115;%%
  %34 citations counted in INSPIRE as of 18 Sep 2013
  
  
  %\cite{Badziak:2013bda}
\bibitem{Badziak:2013bda}
  M.~Badziak, M.~Olechowski and S.~Pokorski,
  ``New Regions in the NMSSM with a 125 GeV Higgs,''
  JHEP {\bf 1306}, 043 (2013)
  [arXiv:1304.5437 [hep-ph]].
  %%CITATION = ARXIV:1304.5437;%%
  %23 citations counted in INSPIRE as of 12 May 2014

%\cite{Badziak:2013gla}
\bibitem{Badziak:2013gla}
  M.~Badziak, M.~Olechowski and S.~Pokorski,
  ``125 GeV Higgs and enhanced diphoton signal of a light singlet-like scalar in NMSSM,''
  arXiv:1310.4518 [hep-ph].
  %%CITATION = ARXIV:1310.4518;%%
  %1 citations counted in INSPIRE as of 12 May 2014
  
  
  %%%%%%%%%%%%%%%%%%%%%%%%%%%%%%%%%%%%%%%%%%%%%%%%%%%%%%%%%%%%%%%
  
  




  
  %\cite{Ross:2011xv}
\bibitem{Ross:2011xv} 
  G.~G.~Ross and K.~Schmidt-Hoberg,
  ``The Fine-Tuning of the Generalised NMSSM,''
Nucl.\ Phys.\ B {\bf 862}, 710 (2012)
[arXiv:1108.1284 [hep-ph]].
%%CITATION = ARXIV:1108.1284;%%
  %52 citations counted in INSPIRE as of 30 Sep 2013

%\cite{Ross:2012nr}
\bibitem{Ross:2012nr} 
  G.~G.~Ross, K.~Schmidt-Hoberg and F.~Staub,
  ``The Generalised NMSSM at One Loop: Fine Tuning and Phenomenology,''
JHEP {\bf 1208}, 074 (2012)
[arXiv:1205.1509 [hep-ph]].
%%CITATION = ARXIV:1205.1509;%%
  %35 citations counted in INSPIRE as of 30 Sep 2013

%\cite{Kaminska:2013mya}
\bibitem{Kaminska:2013mya} 
  A.~Kaminska, G.~G.~Ross and K.~Schmidt-Hoberg,
  ``Non-universal gaugino masses and fine tuning implications for SUSY searches in the MSSM and the GNMSSM,''
arXiv:1308.4168 [hep-ph].
%%CITATION = ARXIV:1308.4168;%%
  %1 citations counted in INSPIRE as of 30 Sep 2013  
  
  
  
%\cite{Dine:2007xi}
\bibitem{Dine:2007xi} 
  M.~Dine, N.~Seiberg and S.~Thomas,
  ``Higgs physics as a window beyond the MSSM (BMSSM),''
Phys.\ Rev.\ D {\bf 76}, 095004 (2007)
[arXiv:0707.0005 [hep-ph]].
%%CITATION = ARXIV:0707.0005;%%
  %135 citations counted in INSPIRE as of 30 Sep 2013

%\cite{Cassel:2009ps}
\bibitem{Cassel:2009ps} 
  S.~Cassel, D.~M.~Ghilencea and G.~G.~Ross,
  ``Fine tuning as an indication of physics beyond the MSSM,''
Nucl.\ Phys.\ B {\bf 825}, 203 (2010)
[arXiv:0903.1115 [hep-ph]].
%%CITATION = ARXIV:0903.1115;%%
  %50 citations counted in INSPIRE as of 30 Sep 2013
  
  %\cite{Carena:2010cs}
\bibitem{Carena:2010cs}
  M.~Carena, E.~Ponton and J.~Zurita,
  ``BMSSM Higgs Bosons at the Tevatron and the LHC,''
  Phys.\ Rev.\ D {\bf 82}, 055025 (2010)
  [arXiv:1005.4887 [hep-ph]].
  %%CITATION = ARXIV:1005.4887;%%
  %27 citations counted in INSPIRE as of 12 May 2014
  
  
  %\cite{Delgado:2010uj}
\bibitem{Delgado:2010uj} 
  A.~Delgado, C.~Kolda, J.~P.~Olson and A.~de la Puente,
  ``Solving the Little Hierarchy Problem with a Singlet and Explicit $\mu$ Terms,''
  Phys.\ Rev.\ Lett.\  {\bf 105}, 091802 (2010)
  [arXiv:1005.1282 [hep-ph]].
  %%CITATION = ARXIV:1005.1282;%%
  %32 citations counted in INSPIRE as of 12 Apr 2013

%\cite{Delgado:2012yd}
\bibitem{Delgado:2012yd} 
  A.~Delgado, C.~Kolda and A.~de la Puente,
  ``Solving the Hierarchy Problem with a Light Singlet and Supersymmetric Mass Terms,''
  Phys.\ Lett.\ B {\bf 710}, 460 (2012)
  [arXiv:1111.4008 [hep-ph]].
  %%CITATION = ARXIV:1111.4008;%%
  %7 citations counted in INSPIRE as of 12 Apr 2013
  
  %\cite{Delgado:2010cw}
\bibitem{Delgado:2010cw} 
  A.~Delgado, C.~Kolda, J.~P.~Olson and A.~de la Puente,
  ``Gauge-mediated embedding of the singlet extension of the minimal supersymmetric standard model,''
  Phys.\ Rev.\ D {\bf 82}, 035006 (2010)
  [arXiv:1005.4901 [hep-ph]].
  %%CITATION = ARXIV:1005.4901;%%
  %7 citations counted in INSPIRE as of 12 Apr 2013
  
  
    
\bibitem{Jeong:2011jk} 
  K.~S.~Jeong, Y.~Shoji and M.~Yamaguchi,
  ``Peccei-Quinn invariant extension of the NMSSM,''
  JHEP {\bf 1204}, 022 (2012)
  [arXiv:1112.1014 [hep-ph]].
  %%CITATION = ARXIV:1112.1014;%%  
 

  
  %\cite{Bergstrom:2000pn}
\bibitem{Bergstrom:2000pn} 
  L.~Bergstrom,
  ``Nonbaryonic dark matter: Observational evidence and detection methods,''
  Rept.\ Prog.\ Phys.\  {\bf 63}, 793 (2000)
  [hep-ph/0002126].
  %%CITATION = HEP-PH/0002126;%%

%\cite{Bertone:2004pz}
\bibitem{Bertone:2004pz} 
  G.~Bertone, D.~Hooper and J.~Silk,
  ``Particle dark matter: Evidence, candidates and constraints,''
  Phys.\ Rept.\  {\bf 405}, 279 (2005)
  [hep-ph/0404175].
  %%CITATION = HEP-PH/0404175;%%
  
  
%\cite{Ade:2013lta}
\bibitem{Ade:2013lta} 
  P.~A.~R.~Ade {\it et al.}  [Planck Collaboration],
  ``Planck 2013 results. XVI. Cosmological parameters,''
  arXiv:1303.5076 [astro-ph.CO].
  %%CITATION = ARXIV:1303.5076;%%
  %43 citations counted in INSPIRE as of 17 Apr 2013
  
  %\cite{Aalseth:2010vx}
\bibitem{Aalseth:2010vx} 
  C.~E.~Aalseth {\it et al.}  [CoGeNT Collaboration],
``Results from a Search for Light-Mass Dark Matter with a P-type Point Contact Germanium Detector,''
  Phys.\ Rev.\ Lett.\  {\bf 106}, 131301 (2011)
  [arXiv:1002.4703 [astro-ph.CO]].
  %%CITATION = ARXIV:1002.4703;%%
  
%\cite{Agnese:2013rvf}
\bibitem{Agnese:2013rvf} 
  R.~Agnese {\it et al.}  [CDMS Collaboration],
  ``Dark Matter Search Results Using the Silicon Detectors of CDMS II,''
  %Submitted to: Phys.Rev.Lett.
  [arXiv:1304.4279 [hep-ex]].
  %%CITATION = ARXIV:1304.4279;%%
  
    
 
    %\cite{Abbiendi:2002qp}
\bibitem{Abbiendi:2002qp} 
  G.~Abbiendi {\it et al.}  [OPAL Collaboration],
  ``Decay mode independent searches for new scalar bosons with the OPAL detector at LEP,''
  Eur.\ Phys.\ J.\ C {\bf 27}, 311 (2003)
  [hep-ex/0206022].
  %%CITATION = HEP-EX/0206022;%%
  %91 citations counted in INSPIRE as of 30 Jul 2013
  
    %\cite{Schael:2006cr}
\bibitem{Schael:2006cr} 
  S.~Schael {\it et al.}  [ALEPH and DELPHI and L3 and OPAL and LEP Working Group for Higgs Boson Searches Collaborations],
  ``Search for neutral MSSM Higgs bosons at LEP,''
  Eur.\ Phys.\ J.\ C {\bf 47}, 547 (2006)
  [hep-ex/0602042].
  %%CITATION = HEP-EX/0602042;%%
  %579 citations counted in INSPIRE as of 12 Apr 2013
  
  %\cite{Abdallah:2004wy}
\bibitem{Abdallah:2004wy} 
  J.~Abdallah {\it et al.}  [DELPHI Collaboration],
  ``Searches for neutral higgs bosons in extended models,''
  Eur.\ Phys.\ J.\ C {\bf 38}, 1 (2004)
  [hep-ex/0410017].
  %%CITATION = HEP-EX/0410017;%%
  %60 citations counted in INSPIRE as of 30 Jul 2013
  

  %\cite{Dermisek:2010mg}
\bibitem{Dermisek:2010mg} 
  R.~Dermisek and J.~F.~Gunion,
  ``New constraints on a light CP-odd Higgs boson and related NMSSM Ideal Higgs Scenarios,''
  Phys.\ Rev.\ D {\bf 81}, 075003 (2010)
  [arXiv:1002.1971 [hep-ph]].
  %%CITATION = ARXIV:1002.1971;%%
  %57 citations counted in INSPIRE as of 18 Jun 2013
  
  %\cite{Andreas:2010ms}
\bibitem{Andreas:2010ms} 
  S.~Andreas, O.~Lebedev, S.~Ramos-Sanchez and A.~Ringwald,
  ``Constraints on a very light CP-odd Higgs of the NMSSM and other axion-like particles,''
  JHEP {\bf 1008}, 003 (2010)
  [arXiv:1005.3978 [hep-ph]].
  %%CITATION = ARXIV:1005.3978;%%
  %18 citations counted in INSPIRE as of 18 Jun 2013
  
     
  %\cite{Barger:2012hv}
\bibitem{Barger:2012hv} 
  V.~Barger, M.~Ishida and W.~-Y.~Keung,
  ``Total Width of 125 GeV Higgs Boson,''
  Phys.\ Rev.\ Lett.\  {\bf 108}, 261801 (2012)
  [arXiv:1203.3456 [hep-ph]].
  %%CITATION = ARXIV:1203.3456;%%
  %39 citations counted in INSPIRE as of 18 Sep 2013
  
  %\cite{Belanger:2013xza}
\bibitem{Belanger:2013xza} 
  G.~Belanger, B.~Dumont, U.~Ellwanger, J.~F.~Gunion and S.~Kraml,
  ``Global fit to Higgs signal strengths and couplings and implications for extended Higgs sectors,''
  arXiv:1306.2941 [hep-ph].
  %%CITATION = ARXIV:1306.2941;%%
  %21 citations counted in INSPIRE as of 18 Sep 2013
  
  
  %\cite{Dobrescu:2012td}
\bibitem{Dobrescu:2012td} 
  B.~A.~Dobrescu and J.~D.~Lykken,
  ``Coupling spans of the Higgs-like boson,''
  JHEP {\bf 1302}, 073 (2013)
  [arXiv:1210.3342 [hep-ph]].
  %%CITATION = ARXIV:1210.3342;%%
  %27 citations counted in INSPIRE as of 18 Sep 2013
  
    %\cite{ALEPH:2005ab}
\bibitem{ALEPH:2005ab} 
  S.~Schael {\it et al.}  [ALEPH and DELPHI and L3 and OPAL and SLD and LEP Electroweak Working Group and SLD Electroweak Group and SLD Heavy Flavour Group Collaborations],
  ``Precision electroweak measurements on the $Z$ resonance,''
  Phys.\ Rept.\  {\bf 427}, 257 (2006)
  [hep-ex/0509008].
  %%CITATION = HEP-EX/0509008;%%
  %808 citations counted in INSPIRE as of 18 Jun 2013
  
    
  %\cite{Abbiendi:2003sc}
\bibitem{Abbiendi:2003sc} 
  G.~Abbiendi {\it et al.}  [OPAL Collaboration],
  ``Search for chargino and neutralino production at s**(1/2) = 192-GeV to 209 GeV at LEP,''
  Eur.\ Phys.\ J.\ C {\bf 35}, 1 (2004)
  [hep-ex/0401026].
  %%CITATION = HEP-EX/0401026;%%
  %100 citations counted in INSPIRE as of 17 Jun 2013
  
  %\cite{Das:2010ww}
\bibitem{Das:2010ww} 
  D.~Das and U.~Ellwanger,
  ``Light dark matter in the NMSSM: upper bounds on direct detection cross sections,''
  JHEP {\bf 1009}, 085 (2010)
  [arXiv:1007.1151 [hep-ph]].
  %%CITATION = ARXIV:1007.1151;%%
  
  %\cite{Jungman:1995df}
\bibitem{Jungman:1995df} 
  G.~Jungman, M.~Kamionkowski and K.~Griest,
  ``Supersymmetric dark matter,''
  Phys.\ Rept.\  {\bf 267}, 195 (1996)
  [hep-ph/9506380].
  %%CITATION = HEP-PH/9506380;%%
  
  %\cite{Greene:1986th}
\bibitem{Greene:1986th} 
  B.~R.~Greene and P.~J.~Miron,
 ``Supersymmetric Cosmology With A Gauge Singlet,''
  Phys.\ Lett.\ B {\bf 168}, 226 (1986).
  %%CITATION = PHLTA,B168,226;%%

%\cite{Flores:1990bt}
\bibitem{Flores:1990bt} 
  R.~Flores, K.~A.~Olive and D.~Thomas,
 ``A New Dark Matter Candidate In The Minimal Extension Of The Supersymmetric Standard Model,''
  Phys.\ Lett.\ B {\bf 245}, 509 (1990).
  %%CITATION = PHLTA,B245,509;%%

%\cite{Olive:1990aj}
\bibitem{Olive:1990aj} 
  K.~A.~Olive and D.~Thomas,
  ``A Light dark matter candidate in an extended supersymmetric model,''
  Nucl.\ Phys.\ B {\bf 355}, 192 (1991).
  %%CITATION = NUPHA,B355,192;%%
  
  %\cite{Abel:1992ts}
\bibitem{Abel:1992ts} 
  S.~A.~Abel, S.~Sarkar and I.~B.~Whittingham,
  ``Neutralino dark matter in a class of unified theories,''
  Nucl.\ Phys.\ B {\bf 392}, 83 (1993)
  [hep-ph/9209292].
  %%CITATION = HEP-PH/9209292;%%
  
%\cite{Stephan:1997rv}
\bibitem{Stephan:1997rv} 
  A.~Stephan,
  ``Dark matter constraints on the parameter space and particle spectra in the nonminimal SUSY standard model,''
  Phys.\ Lett.\ B {\bf 411}, 97 (1997)
  [hep-ph/9704232].
  %%CITATION = HEP-PH/9704232;%%
  
  %\cite{Gunion:2005rw}
\bibitem{Gunion:2005rw} 
  J.~F.~Gunion, D.~Hooper and B.~McElrath,
  ``Light neutralino dark matter in the NMSSM,''
  Phys.\ Rev.\ D {\bf 73}, 015011 (2006)
  [hep-ph/0509024].
  %%CITATION = HEP-PH/0509024;%%


  
  %\cite{Draper:2010ew}
\bibitem{Draper:2010ew} 
  P.~Draper, T.~Liu, C.~E.~M.~Wagner, L.~-T.~Wang and H.~Zhang,
  ``Dark Light Higgs,''
  Phys.\ Rev.\ Lett.\  {\bf 106}, 121805 (2011)
  [arXiv:1009.3963 [hep-ph]].
  %%CITATION = ARXIV:1009.3963;%%
  %47 citations counted in INSPIRE as of 26 Sep 2013

%\cite{Carena:2011jy}
\bibitem{Carena:2011jy} 
  M.~Carena, N.~R.~Shah and C.~E.~M.~Wagner,
  ``Light Dark Matter and the Electroweak Phase Transition in the NMSSM,''
  Phys.\ Rev.\ D {\bf 85}, 036003 (2012)
  [arXiv:1110.4378 [hep-ph]].
  %%CITATION = ARXIV:1110.4378;%%
  %16 citations counted in INSPIRE as of 26 Sep 2013
  
  %\cite{Kozaczuk:2013spa}
\bibitem{Kozaczuk:2013spa} 
  J.~Kozaczuk and S.~Profumo,
  ``Light NMSSM Neutralino Dark Matter in the Wake of CDMS II and a 126 GeV Higgs,''
  arXiv:1308.5705 [hep-ph].
  %%CITATION = ARXIV:1308.5705;%%
  
  
  %\cite{Kolb:1990vq}
\bibitem{Kolb:1990vq} 
  E.~W.~Kolb and M.~S.~Turner,
  ``The Early universe,''
  Front.\ Phys.\  {\bf 69}, 1 (1990).
  %%CITATION = FRPHA,69,1;%%
  %228 citations counted in INSPIRE as of 12 Apr 2013
  
   %\cite{Nihei:2001qs}
\bibitem{Nihei:2001qs} 
  T.~Nihei, L.~Roszkowski and R.~Ruiz de Austri,
  ``Towards an accurate calculation of the neutralino relic density,''
  JHEP {\bf 0105}, 063 (2001)
  [hep-ph/0102308].
  %%CITATION = HEP-PH/0102308;%% 
  
  %\cite{Nihei:2002ij}
\bibitem{Nihei:2002ij} 
  T.~Nihei, L.~Roszkowski and R.~Ruiz de Austri,
 ``Exact cross-sections for the neutralino WIMP pair annihilation,''
  JHEP {\bf 0203}, 031 (2002)
  [hep-ph/0202009].
  %%CITATION = HEP-PH/0202009;%%
  
  %\cite{Bauer:2013ihz}
\bibitem{Bauer:2013ihz} 
  D.~Bauer, J.~Buckley, M.~Cahill-Rowley, R.~Cotta, A.~Drlica-Wagner, J.~Feng, S.~Funk and J.~Hewett {\it et al.},
  ``Dark Matter in the Coming Decade: Complementary Paths to Discovery and Beyond,''
  arXiv:1305.1605 [hep-ph].
  %%CITATION = ARXIV:1305.1605;%%

  
 %\cite{Gutlein:2010tq}
\bibitem{Gutlein:2010tq} 
  A.~Gutlein, C.~Ciemniak, F.~von Feilitzsch, N.~Haag, M.~Hofmann, C.~Isaila, T.~Lachenmaier and J.~-C.~Lanfranchi {\it et al.},
  ``Solar and atmospheric neutrinos: Background sources for the direct dark matter search,''
  Astropart.\ Phys.\  {\bf 34}, 90 (2010)
  [arXiv:1003.5530 [hep-ph]].
  %%CITATION = ARXIV:1003.5530;%%
  
  
 %\cite{Akerib:2013tjd}
\bibitem{Akerib:2013tjd} 
  D.~S.~Akerib {\it et al.}  [LUX Collaboration],
  ``First results from the LUX dark matter experiment at the Sanford Underground Research Facility,''
  arXiv:1310.8214 [astro-ph.CO].
  %%CITATION = ARXIV:1310.8214;%%
  
  
  
  %%%%%%%%%%%           Additional reference        %%%%%%%%%
  
  
  
%%\cite{Hooper:2012sr}
%\bibitem{Hooper:2012sr} 
%  D.~Hooper, C.~Kelso and F.~S.~Queiroz,
%  ``Stringent and Robust Constraints on the Dark Matter Annihilation Cross Section From the Region of the Galactic Center,''
%  Astropart.\ Phys.\  {\bf 46}, 55 (2013)
%  [arXiv:1209.3015 [astro-ph.HE]].
%  %%CITATION = ARXIV:1209.3015;%%



\end{thebibliography}
\end{document}